\documentclass[journal,11pt,draftclsnofoot,onecolumn]{IEEEtran}

\usepackage{cite}
\usepackage{algorithm}
\usepackage{algpseudocode}
\usepackage[small,bf]{caption}
\usepackage[cmex10]{amsmath}
\usepackage{amssymb,latexsym,epsfig,subfigure,epic,amscd,mathrsfs,euscript,eufrak}
\usepackage{color} 

\newtheorem{myprop}{\bf{Proposition}}

\newcommand{\argmin}{\operatornamewithlimits{arg\,min}}
\newcommand{\bbR}{\mathbb{R}}

\DeclareMathOperator{\tr}{tr}
\DeclareMathOperator{\card}{card}

\DeclareMathOperator*{\minimize}{\text{minimize}}
\DeclareMathOperator*{\st}{\text{subject to}}
\DeclareMathAlphabet\mathbfcal{OMS}{cmsy}{b}{n}
\def\NoNumber#1{{\def\alglinenumber##1{}\State #1}\addtocounter{ALG@line}{-1}}
\begin{document}

\title{Optimal Periodic Sensor Scheduling in\\
Networks of Dynamical Systems}

\author{Sijia~Liu,~\IEEEmembership{Student Member,~IEEE,}
        Makan~Fardad,~\IEEEmembership{Member,~IEEE,}
        Engin~Masazade,~\IEEEmembership{Member,~IEEE,}
        and~Pramod~K.~Varshney,~\IEEEmembership{Fellow,~IEEE}
\thanks{S. Liu, M. Fardad and P. K. Varshney are with the Department
of Electrical Engineering and Computer Science, Syracuse University, Syracuse,
NY, 13244 USA e-mail: \{sliu17, makan, varshney\}@syr.edu.}
\thanks{E. Masazade is with the Department of Electrical and Electronics Engineering, Yeditepe University, Istanbul, 34755, Turkey e-mail: engin.masazade@yeditepe.edu.tr.}
\thanks{The work of S. Liu and P. K. Varshney was supported by the U.S. Air Force Office of Scientific Research (AFOSR) under Grants FA9550-10-1-0263 and FA9550-10-1-0458. The work of M. Fardad was supported by the National Science Foundation under award CMMI-0927509. The work of E. Masazade was supported by the Scientic and Technological Research Council of Turkey (TUBITAK) under Grant 113E220.}
}


%



\maketitle

\begin{abstract}
We consider the problem of finding optimal time-periodic sensor
schedules for estimating the state of discrete-time dynamical systems.
We assume that {multiple} sensors have been deployed and 
that the sensors are subject to resource constraints, which limits
the number of times each can be activated over one period of the periodic schedule. 
We seek an algorithm that strikes a balance between estimation 
accuracy and total sensor activations over one period. We make
a correspondence between active sensors and the nonzero columns of estimator gain. We formulate an optimization problem in which
we minimize the trace of the error covariance with respect to the estimator gain
while simultaneously penalizing the number of nonzero columns of the estimator gain.
This optimization problem is combinatorial in nature, and we employ 
the alternating direction method of multipliers (ADMM) to find its
locally optimal solutions. Numerical results and comparisons with other sensor scheduling algorithms in the literature are provided to 
illustrate the effectiveness of our proposed method. 
\end{abstract}

\begin{IEEEkeywords}
Dynamical systems,
{alternating direction method of multipliers,}
state estimation,
sensor networks,
sensor scheduling,
sparsity.
\end{IEEEkeywords}

\IEEEpeerreviewmaketitle

\section{Introduction}


\IEEEPARstart{W}{ireless} sensor networks, consisting of a large number of spatially distributed sensors, have been used in a wide range of application areas such as 
environment monitoring, source localization and object tracking \cite{RUR2004,ERP2010, MRV2012_j}.
In a given region of interest, sensors observe the unknown state (e.g., field intensity or target location) which commonly evolves as part of a linear dynamical system. A fusion center receives all the measurements and estimates the state over the entire spatial domain.
However, due to the constraints on communication bandwidth and sensor battery life, it may not be desirable to have all the sensors report their measurements at all time instants. Therefore, the problem of sensor selection/scheduling arises, which seeks to activate different subsets of sensors at different time instants in order to attain an optimal tradeoff between estimation accuracy and energy use. 

Over the last decade, 
sensor selection/scheduling problems for state estimation of linear systems have been extensively studied in the literature \cite{EFJP03,HPYE04,Gupta04,SS2009,YRB2011,shevar13,CMP04,JPF03,PVJ2007,Tomlin2012,WMJ2010,Jiming2011,TM2011}, 
where several variations of the problem have been addressed according to the types of cost functions, time horizons, heuristic algorithms, and energy and topology constraints. Many research efforts have focused on myopic sensor scheduling \cite{EFJP03,HPYE04,Gupta04,SS2009}, where at every instant the search is for the best sensors to be activated at the next time step (as opposed to a longer time horizon). 
However, myopic selection strategies get trapped in local optima and perform poorly in some cases, such as sensor networks with sensing holes \cite{JPF03}. 
But if the length of time horizon becomes large or infinite then
finding an optimal non-myopic schedule is difficult,
because the number of sensor sequences grows prohibitively large as the time horizon grows. 
Therefore, some researchers have considered the problem of periodic sensor schedules on an infinite time horizon \cite{WMJ2010,Jiming2011,JEM11,TM2011,EC90}. 

In \cite{PVJ2007,Tomlin2012}, periodicity in the optimal sensor schedule was observed even 
for finite time horizon problems in which a periodic schedule was not assumed {\em a priori}.
A sufficient condition for the existence of periodicity for the sensor scheduling problem over an infinite time horizon was first suggested in \cite{EC90}.
Furthermore, in \cite{WMJ2010} it was proved that the optimal sensor schedule for an infinite horizon problem can be approximated arbitrarily well by a periodic schedule with a finite period. 
We emphasize that the results in \cite{WMJ2010} are nonconstructive, in the sense that it is shown that the optimal sensor schedule is time-periodic but an algorithm for obtaining this schedule, or even the length of its period, is not provided. Although periodicity makes infinite horizon sensor scheduling problems tractable via the design of an optimal schedule over a finite period, it {poses} other challenges in problem formulation and optimization compared to  conventional sensor scheduling. 

{
In this paper, we seek a general framework to design optimal periodic sensor schedules subject to measurement frequency constraints. Measurement frequency constraints imply that each sensor has a bound on the number of times it can be active over a time period of length $K$. Similar constraints have been considered in \cite{YRB2011,PVJ2007,HJB2009} and referred to as energy constraints, and transmission or communication bounds. 
To achieve our goal, we seek an optimal dynamic estimator, in the form of a 
time-periodic Kalman filter, that also respects the measurement frequency constraints. This can be interpreted as a design problem in which {\em both} the sensor activation schedules, \textit{and} the estimator gains used to combine the sensor measurements, are {\em jointly optimized}. To allow for additional design flexibility, we introduce into the optimization formulation sparsity-promoting penalty functions that encourage fewer measurements at every time instant of the periodic horizon. This can be used to generate arbitrarily sparse sensor schedules that employ a minimal number of active sensors.
}


The design of optimal periodic sensor schedules has been recently studied in \cite{Jiming2011,JEM11,TM2011}. In \cite{Jiming2011}, the authors construct the optimal periodic schedule only for two sensors. 
For a multiple sensor scenario, the work of \cite{TM2011} studied the problem of periodic sensor scheduling by assuming the process noise to be very small, which results in a linear matrix inequality (LMI) problem. As a consequence of the assumption that the process noise is negligible, the ordering of the measurements does not factor into the solution of this LMI problem.  
Clearly, a sensor schedule in which the order of sensor activations is irrelevant can not be optimal for some sensor scheduling problems. For example, it was shown in \cite{NVMn2005} that temporally staggered sensor schedules constitute the optimal sensing policy. 
In \cite{JEM11}, a lower bound on the performance of scheduling sensors over an infinite time horizon is obtained, and then an open-loop periodic switching policy is constructed by using a doubly substochastic matrix. The authors show that the presented switching policy 
achieves the best estimation performance as 
the period length goes to zero (and thus sensors are switched as fast as possible). 
In this paper, a comparison of both the performance and the computational complexity of our methodology with the existing work in \cite{Jiming2011,JEM11,TM2011} will be provided.

The sensor scheduling framework presented in this paper relies on making a one-to-one correspondence between every sensor and a column of the estimator gain. 
Namely, a sensor being off at a certain time instant is equivalent to the corresponding column of the estimator gain being identically zero. This idea has been exploited in our earlier work \cite{EMP2012} on sparsity-promoting extended Kalman filtering, where sensors are scheduled only for the next time step and have no resources constraints involved. Different from \cite{EMP2012}, we consider a periodic sensor scheduling problem on an infinite time horizon, where measurement frequency constraints and periodicity place further restrictions on the number of nonzero columns of the time-periodic Kalman filter gain matrices.

Counting and penalizing {the number of} nonzero columns of the estimator gain, which in this work is performed via the use of the cardinality function, results in combinatorial optimization problems that are intractable in general. It has been recently observed in \cite{FMM2013J,FMM2011,EMP2012} that the alternating direction method of multipliers (ADMM) is a powerful tool for solving optimization problems that include cardinality functions. Particularly, reference \cite{FMM2013J} considers the problem of finding optimal sparse state feedback gains and demonstrates the effectiveness of ADMM in finding such gains.
However, different from \cite{FMM2013J}, we extend the application of ADMM to
account for the time periodicity. Furthermore, we incorporate measurement frequency constraints where subtle relationships between the sparsity-promoting parameter and the frequency parameter come into play.

The main contributions of this paper can be summarized as follows.
\begin{itemize}
\item 
We develop a general optimization framework for the joint design of optimal periodic sensor schedules (on an infinite time horizon) and optimal estimator (Kalman filter) gain matrices.
\item 
We demonstrate that the optimal periodic Kalman filter gain matrices should satisfy a coupled sequence of periodic Lyapunov recursions. We introduce a new block-cyclic representation to transform the coupled matrix recursions into algebraic matrix equations. In particular, this allows the application of the efficient Anderson-Moore method in solving the optimization problem.
\item 
Through application of the alternating direction method of multipliers, we uncover subtle relationships between the frequency constraint parameter, the sparsity-promoting parameter, and the sensor schedule.
\item 
We present a comparison of both the performance and the computational complexity of our methodology with other prominent work in the literature. We demonstrate that our method performs as well or significantly better than these works, and is computationally efficient for sensor scheduling in problems with large-scale dynamical systems.
\end{itemize}

The rest of the paper is organized as follows. In Section \ref{sec: motivation}, we motivate the problem of periodic sensor scheduling on an infinite time horizon. In Section \ref{sec: problem_form}, we formulate the sparsity-promoting periodic sensor scheduling problem. In Section \ref{sec: ADMM}, we invoke the ADMM method, which leads to a pair of efficiently solvable subproblems. In Section \ref{sec: sim}, we illustrate the effectiveness of our proposed approach through examples. Finally, in Section \ref{sec:conclusion} we summarize our work and discusses future research directions.


\section{Periodicity of infinite horizon sensor scheduling}
\label{sec: motivation}

Consider a discrete-time linear dynamical system evolving according to the equations 
\begin{align}
{\mathbf x}_{k+1} &= {\mathbf A}{\mathbf x}_k + \mathbf B {\mathbf w}_k, \label{eq: state}\\
\mathbf y_k &= \mathbf C \mathbf x_k + \mathbf v_k, \label{eq: meas}
\end{align}
where $\mathbf x_k \in \bbR^N$ is the state vector at time $k$, $\mathbf y_k \in \bbR^M$ is 
the measurement vector whose $m$th entry corresponds to a scalar observation from sensor $m$, $\mathbf A$, $\mathbf B$, and $\mathbf C$ are matrices of appropriate dimensions. 
The inputs $\mathbf w_k$ and $\mathbf v_k$ are white, Gaussian, zero-mean random vectors with covariance matrices $\mathbf Q$ and $\mathbf R$, respectively. Finally, we assume that 
$(\mathbf A, \mathbf C)$ is detectable and $(\mathbf A, \boldsymbol \Sigma)$ is stabilizable, where $\boldsymbol \Sigma \boldsymbol \Sigma^T =\mathbf B \mathbf Q \mathbf B^T$.

For ease of describing the sensor schedule, we introduce the auxiliary binary variables $\zeta_{k,m} \in \{0,1\}$, to represent whether or not the $m$th sensor is activated at time $k$. 
The sensor schedule over an infinite time horizon can then be denoted by 
$\boldsymbol \mu_{\infty} = [\boldsymbol \zeta_{1}, \boldsymbol \zeta_{2},\ldots]$, 
where the vector $\boldsymbol \zeta_{k} = [\zeta_{k,1}, \ldots, \zeta_{k,M}]^T$ indicates which sensors are active at time $k$.
The performance of an infinite-horizon sensor schedule is then measured as follows
\cite{Jiming2011,TM2011}, 
\begin{align}
J (\boldsymbol \mu_{\infty}) \triangleq \lim_{\overline{K} \rightarrow \infty} \frac{1}{\overline{K}} \sum_{k=1}^{\overline{K}} \mathrm{tr}\left( \mathbf P_k \right)
\label{eq: J_inf}
\end{align}
where $\mathbf P_k$ is the estimation error covariance at time $k$ under the sensor schedule $\boldsymbol \mu_{\infty}$. Due to the combinatorial nature of the problem, it is intractable to find the optimal sensor schedule that minimizes the cost (\ref{eq: J_inf}) in general\cite{TM2011}.

In \cite{EC90}, it was suggested that the optimal sensor schedule can be treated as a time-periodic schedule over the infinite time horizon if the system (\ref{eq: state})-(\ref{eq: meas}) is detectable and stabilizable. Furthermore, in \cite{WMJ2010} it was proved that the optimal sensor schedule for an infinite horizon problem can be approximated arbitrarily well by a periodic schedule with a finite period, and that the error covariance matrix converges to a unique limit cycle. In this case, the cost in (\ref{eq: J_inf}) can be rewritten as 
\begin{align}
J ( \boldsymbol \mu_{K} ) =  \frac{1}{K} \sum_{k=0}^{K-1} \mathrm{tr}\left( \mathbf P_k \right)
\label{eq: J_periodic}
\end{align}
where $K$ is the length of the period and $\mathbf P_k$ is the error covariance matrix at instant $k$ of its limit cycle. In this work, similar to \cite{WMJ2010,Jiming2011,TM2011}, we assume the length $K$ of the period is given. To the best knowledge of the authors, it is still an open problem to find the optimal period length.

\section{Problem formulation}
\label{sec: problem_form}

For the discrete-time linear dynamical system (\ref{eq: state})--(\ref{eq: meas}), we consider state estimators of the form 
\begin{align*}
{\hat {\mathbf x} }_{k+1} 
&=  \mathbf A   {\hat {\mathbf x}}_{k} + \mathbf L_k(  \mathbf y_k -  \mathbf C \hat {\mathbf x}_{k} ) 
= ( \mathbf A -  \mathbf L_k  \mathbf C) \hat{\mathbf x}_{k} +  \mathbf L_k  \mathbf y_k,
\end{align*} 
where $\mathbf L_k$ is the estimator gain (also known as the observer gain \cite{Kalman}) at time $k$. In what follows we aim to determine the matrices $\mathbf L_k$, $k=0,1,\ldots$, by solving an optimization problem that, in particular, promotes the column sparsity of $\mathbf L_k$. We define the estimation error covariance $\mathbf P_k$ as
\begin{equation*}
\mathbf P_{k} = \mathcal E \{ (\mathbf{x}_k -  \hat{\mathbf x}_k )(\mathbf {x}_k - \hat{\mathbf x}_k )^T\}, 
\end{equation*}
where $\mathcal E$ is the expectation operator.%
\footnote{In the system theory literature, $\hat{\mathbf x}_k$ and $\mathbf P_k$ are often denoted by $\hat {\mathbf x}_{k|k-1}$ and $\mathbf P_{k|k-1}$; here we use $\hat {\mathbf x}_{k}$ and $\mathbf P_k$ for simplicity of notation.} 
It is easy to show that $\mathbf P_k$ satisfies the Lyapunov recursion
\begin{equation}
\mathbf P_{k+1} = (\mathbf A- \mathbf L_k \mathbf C) \mathbf P_k 
(\mathbf A - \mathbf L_k \mathbf C)^T + \mathbf B \mathbf Q \mathbf B^T + \mathbf L_k \mathbf R \mathbf L_k^T.
\label{eq: P_lyapunov}
\end{equation}
Finally, partitioning the matrices $\mathbf L_k$ and $\mathbf C$ into their respective columns and rows, we have
\begin{eqnarray}
\mathbf L_k \mathbf C =& \!\!\!\!\!\!\!\!\!
\begin{bmatrix}
      \mathbf L_{k,1} & \mathbf L_{k,2} & \ldots & \mathbf L_{k,M}
     \end{bmatrix}
\begin{bmatrix}
      \mathbf C_1^T \\
      \mathbf C_2^T \\
      \vdots \\
      \mathbf C_M^T
     \end{bmatrix} \nonumber
     \\
=& \mathbf L_{k,1} \mathbf C_1^T + \mathbf L_{k,2} \mathbf C_2^T + \cdots + \mathbf L_{k,M} \mathbf C_M^T,
\label{eq: partition}
\end{eqnarray}
where we assume that each row of $\mathbf C$ characterizes the measurement of one sensor. Therefore, each column of the matrix $\mathbf L_k$ can be thought of as corresponding to the measurement of a particular sensor.

In estimation and inference problems using wireless sensor networks, minimizing the energy consumption of sensors is often desired \cite{HJB2009, YRB2011}. Therefore, we seek algorithms that schedule the turning on and off of the sensors in order to strike a balance between energy consumption and estimation performance. 
Suppose, for example, that at time step $k$ only the $\nu$th sensor reports a measurement. In this case, it follows from (\ref{eq: partition}) that $\mathbf L_k \mathbf C = \mathbf L_{k, \nu} \mathbf C_{\nu}^T$, where $\mathbf C_{\nu}^T$ is the $\nu$th row of $\mathbf C$. This can also be interpreted as having the column vectors $\mathbf L_{k,m}$ equal to zero for all $m \neq \nu$. Thus, hereafter we assume that the measurement matrix $\mathbf C$ is constant and the scheduling of the sensors is captured by the nonzero columns of the estimator gains $\mathbf L_k$, in the sense that if $\mathbf L_{k,m}=\boldsymbol{0}$ then at time $k$ the $m$th sensor is not making a measurement.


As stated in Sec.\,\ref{sec: motivation}, 
in this work we search for optimal {\em time-periodic} sensor schedules, i.e., we seek optimal sequences $\{ \mathbf L_k\}_{k = 0, 1, \ldots, K-1}$ and $\{\mathbf P_k \}_{k = 0,1, \ldots, K-1}$ that satisfy
\begin{align}
\mathbf L_{k+K} = \mathbf L_k,
~~~
\mathbf P_{k+K} = \mathbf P_k,
\label{eq: period}
\end{align} where $K$ is a given period. Note that the choice of $K$ is not a part of the optimization problem considered in this paper. 
As suggested in \cite{WMJ2010}, one possible procedure for choosing $K$ is to find the optimal sensor schedule for gradually-increasing values of $K$ until the performance ceases to improve significantly. 
Furthermore, the condition on the periodicity of $\mathbf P_k$ assumes that the system and estimator with $\mathbf L_{k+K}=\mathbf L_k$ have been running for a long time so that $\mathbf P_k$ has reached its steady-state limit cycle \cite{WMJ2010}. In this paper, we consider $k= - \infty $ as the initial time and without loss of generality consider the design of $\mathbf L_k$ over the period $k = 0, 1, \ldots, K-1 $, {when the system has statistically settled into its periodic cycle.}

To incorporate the energy constraints on individual sensors over a period of length $K$, we consider 
\begin{align}
\displaystyle{\sum_{k=0}^{K-1}} \card\! \big(   \|  \mathbf L_{k,m} \|_2  \big) \leq {\eta_m},
~~
m = 1, 2, \ldots, M,
\label{eq: freq_cons}
\end{align}
where $\eta_m$ denotes the {measurement frequency bound}. This implies that the $m$th sensor can make and transmit at most $\eta_m$ measurements over the period of length $K$. For simplicity, we assume $\eta_1 = \eta_2 = \ldots = \eta_M = \eta$. We remark that the proposed sensor scheduling methodology in this article applies equally well to the case where the $\eta_i$ are not necessarily equal to each other.

Next, we formulate the optimal periodic sensor scheduling problem considered in this work, and then elaborate on the details of our formulation. We pose the optimal sensor scheduling problem as the optimization problem
\begin{eqnarray}
\hspace*{-0.13in}
\begin{array}{ll}
\text{minimize}
\hspace*{-0.12in}& 
\displaystyle{\sum_{k=0}^{K-1}} \tr(\mathbf P_k) + \gamma \displaystyle{\sum_{k=0}^{K-1}} g(\mathbf L_k)
\\[0.15cm]
\text{subject to}\hspace*{-0.12in}
&
\left \{ 
\begin{array}{l}
\hspace*{-0.08in}\text{Lyapunov recursion (\ref{eq: P_lyapunov}) for $k \hspace*{-0.03in} = \hspace*{-0.01in} 0, \hspace*{-0.01in} 1, \hspace*{-0.01in} \ldots, \hspace*{-0.01in} K\!\!-\!\!1$,} \\ [0.15cm]
\hspace*{-0.08in}\text{periodicity condition (\ref{eq: period}),} \\ [0.15cm]
\hspace*{-0.08in}\text{measurement frequency constraints (\ref{eq: freq_cons}),}
\end{array}


\right.
\end{array}
\hspace{-0.35in}
\label{eq: obj_1}
\end{eqnarray}
where the matrices $\{ \mathbf L_k \}_{k=0,\cdots,K-1}$ are the optimization variables, $\card (\cdot)$ denotes the cardinality function which gives the number of nonzero elements of its (vector) argument, and 
\begin{align}
\hspace*{-0.05in}
g(\mathbf L_k) \hspace*{-0.02in} :=&\,  \card\! \big( \begin{bmatrix}
      \| \mathbf L_{k,1} \|_2 & \| \mathbf L_{k,2} \|_2 & \cdots  & \| \mathbf L_{k,M} \|_2
     \end{bmatrix} \big).
\label{eq: card_g0}
\end{align} 
Therefore $g(\mathbf L_k)$ is equal to the number of nonzero columns of $\mathbf L_k$, also referred to as the column-cardinality of $\mathbf L_k$. The incorporation of the sparsity-promoting term $g(\cdot)$ in the objective function encourages the use of a small subset of sensors at each time instant. The positive scalar $\gamma$ characterizes the relative importance of the two conflicting terms in the objective, namely the relative importance of achieving good estimation performance versus activating a small number of sensors. 

Note that (\ref{eq: obj_1}) is a combinatorial problem \cite{Boyd2004_bk} and, for large systems, computationally intractable in general. Motivated by \cite{FMM2013J}, in the next section we employ the alternating direction method of multipliers (ADMM) to solve (\ref{eq: obj_1}). We demonstrate that the application of ADMM leads to a pair of efficiently solvable subproblems.


\section{Optimal Periodic Sensor Scheduling using ADMM}
\label{sec: ADMM}

In this section, we apply ADMM to the sensor scheduling problem (\ref{eq: obj_1}). Our treatment uses ideas introduced in \cite{FMM2013J}, where ADMM was used for the identification of optimal sparse state-feedback gains. We extend the framework of \cite{FMM2013J} to account for the time periodicity of the estimator gains, their sparsity across both space and time, and the addition of measurement frequency constraints on individual sensors.

We begin by reformulating the optimization problem in (\ref{eq: obj_1}) in a way that lends itself to the application of ADMM. For $\mathbf P_k$ that satisfies the Lyapunov recursion in (\ref{eq: P_lyapunov}), it is easy to show that
\begin{align*}
\mathbf P_k &= \mathbf B \mathbf Q \mathbf B^T + \mathbf L_{k-1} \mathbf R \mathbf L_{k-1}^T \\
&~~ + \sum_{n = k-1}^{- \infty} (\mathbf A- \mathbf L_{k-1} \mathbf C) \cdots (\mathbf A- \mathbf L_n \mathbf C) \\
&\hspace{0.6in} \cdot (\mathbf B \mathbf Q \mathbf B^T + \mathbf L_{n-1} \mathbf R \mathbf L_{n-1}^T) \\
&\hspace{0.6in} \cdot (\mathbf A - \mathbf L_n \mathbf C)^T \cdots (\mathbf A - \mathbf L_{k-1} \mathbf C)^T.
\end{align*}
Invoking the periodicity of $\mathbf L_k$, $\tr(\mathbf P_k)$ can be expressed as a function $f_k$ of $\{ \mathbf L_k \}_{k=0, \cdots, K-1}$ so that the optimization problem (\ref{eq: obj_1}) can be rewritten as
\[
\begin{array}{ll}
\text{minimize}
&
\displaystyle{\sum_{k=0}^{K-1}} f_k (\mathbf L_0, \cdots, \mathbf L_{K-1}) + \gamma \displaystyle{\sum_{k=0}^{K-1}} g(\mathbf L_k)
\\[0.15cm]
\text{subject to}
&
\displaystyle{\sum_{k=0}^{K-1}} \card\! \big(   \| \mathbf L_{k,m} \|_2  \big) \leq {\eta},
~~
m = 1, 2, \ldots, M.
\end{array}
\hspace{-0.35in}
\]
We next introduce the indicator function corresponding to the constraint set of the above optimization problem as \cite{FMM2011}
\begin{equation}
\mathcal{I}(\{ \mathbf L_k \}) = \left\{
  \begin{array}{l l}
    0 &  \text{if $\sum_{k=0}^{K-1} \card\! \big(  \| \mathbf  L_{k,m} \|_2  \big) \leq \eta$}\\
      &  \text{for $m=1, 2, \ldots, M$}, \\[0.2cm]
    +\infty & \text{otherwise},\\
  \end{array} \right.
\label{eq: indicator}
\end{equation}
where for notational simplicity we have used, and henceforth will continue to use, $\{ \cdot \}$ instead of $\{ \cdot \}_{k =0,\ldots, K-1}$.
Incorporating the indicator function into the objective function, problem (\ref{eq: obj_1}) is equivalent to the unconstrained optimization problem
\[
\begin{array}{ll}
\minimize
&
\displaystyle{\sum_{k=0}^{K-1}} f_k (\{ \mathbf L_k\}) + \gamma \displaystyle{\sum_{k=0}^{K-1}} g(\mathbf L_k) + \mathcal{I}(\{ \mathbf L_k \}).
\end{array}
\]
Finally, we introduce the new set of variables $\{ \mathbf G_k \}$, together with the new set of constraints $\mathbf L_k = \mathbf G_k$, $k=0, 1, \ldots, K-1$, and formulate
\begin{equation}
\begin{array}{ll}
\minimize
&
\displaystyle{\sum_{k=0}^{K-1}} f_k (\{ \mathbf L_k\}) + \gamma \displaystyle{\sum_{k=0}^{K-1}} g(\mathbf G_k) + \mathcal{I}(\{ \mathbf G_k \})  \\
\st
& \mathbf L_k = \mathbf G_k, \quad  k=0, 1, \ldots, K-1,
\end{array}
\label{eq: obj_1_admm}
\end{equation}
which is now in a form suitable for the application of ADMM.

The augmented Lagrangian \cite{SNEn2011,FMM2013J} corresponding to optimization problem (\ref{eq: obj_1_admm}) is given by
\begin{align}
&\mathscr{L}(\{\mathbf L_k\},\{\mathbf G_k\},\{\boldsymbol \Lambda_k\}) \nonumber\\
&~~ = \sum_{k=0}^{K-1} f_k(\{\mathbf L_k\}) + \gamma \sum_{k=0}^{K-1} g(\mathbf G_k) + \mathcal{I}(\{ \mathbf G_k \}) \nonumber\\
&~~~~ + \sum_{k=0}^{K-1}\tr[\boldsymbol \Lambda_k (\mathbf L_k - \mathbf G_k)] + \frac{\rho}{2} \sum_{k=0}^{K-1}||\mathbf L_k - \mathbf G_k||_F^2,
\label{eq: ALag1}
\end{align}
where the matrices $\{\boldsymbol \Lambda_k \}$ are the Lagrange multipliers (also referred to as the dual variables), the scalar $\rho > 0$ is a penalty weight, and $\| \cdot \|_F$ denotes the Frobenius norm of a matrix, $\| \mathbf X \|_F^2 = \tr(\mathbf X^T \mathbf X)$.
The ADMM algorithm can be described as follows \cite{SNEn2011}. For $i=0,1,\ldots$, we iteratively execute the following three steps
\begin{align}
\hspace*{-0.05in}\{\mathbf L_k^{i+1}\}
&:= \argmin_{\{\mathbf L_k\}} ~ \mathscr{L} (\{\mathbf L_k\},\{\mathbf G_k^i\},\{\boldsymbol \Lambda_k^i\}),
\label{eq: L1} \\
\hspace*{-0.05in} \{\mathbf G_k^{i+1}\}
&:= \argmin_{\{\mathbf G_k\}} ~ \mathscr{L} (\{\mathbf L_k^{i+1}\},\{\mathbf G_k\},\{\boldsymbol \Lambda_k^i\}),
\label{eq: G1}\\
\hspace*{-0.05in} \boldsymbol \Lambda_k^{i+1}
&:= \boldsymbol \Lambda_k^i + \rho (\mathbf L_k^{i+1}- \mathbf G_k^{i+1}), k = 0, 1, \ldots, K\!-\!1,
\label{eq: Dual1}
\end{align}
until both of the conditions
$
\sum_{k=0}^{K-1}\|\mathbf L_k^{i+1}- \mathbf G_k^{i+1}\|_F \leq \epsilon,
$
and 
$\sum_{k=0}^{K-1}\|\mathbf G_k^{i+1}- \mathbf G_k^{i}\|_F \leq \epsilon
$
are satisfied.

The rationale behind using ADMM can be described as follows \cite{FMM2013J}. The original nonconvex optimization problem (\ref{eq: obj_1}) is difficult to solve due to the nondifferentiability of the sparsity-promoting function $g$. By defining the new set of variables $\{\mathbf G_k\}$, we effectively separate the original problem into an ``$\mathbf L$-minimization'' step (\ref{eq: L1}) and a ``$\mathbf G$-minimization" step (\ref{eq: G1}), of which the former can be addressed using variational methods and descent algorithms and the latter can be solved analytically. 

We summarize our proposed method on periodic sensor scheduling in Algorithm\,1. In the subsections that follow, we will elaborate on each of the steps involved in the implementation of Algorithm\,1 and the execution of the minimization problems (\ref{eq: L1}) and (\ref{eq: G1}).

\begin{algorithm}
\caption{ADMM-based sensor scheduling algorithm}
\begin{algorithmic}[1]
\State \textbf{Require:} Choose $\rho$, $\epsilon$. Initialize ADMM using $\{\boldsymbol \Lambda_k^0\} = \{\mathbf G_k^0\} = \{\mathbf 0\}$ and $\{\mathbf L_k^0\}$ from (\ref{eq: iniL}).
\For{$i=0,1,\ldots$}
\State Obtain $\{\mathbf L_k^{i+1}\}$ using Algorithms\,2-3.
\State Obtain $\{\mathbf G_k^{i+1}\}$ using Algorithm\,4.
\State Obtain $\{\boldsymbol \Lambda_k^{i+1}\}$ using $\boldsymbol \Lambda_k^{i+1}
= \boldsymbol \Lambda_k^i + \rho (\mathbf L_k^{i+1}-\mathbf G_k^{i+1})$, 
\hspace*{0.15in}
$k = 0,1,\ldots, K-1$.
\State \textbf{until} 
$\sum_{k=0}^{K-1}\|\mathbf L_k^{i+1}-\mathbf G_k^{i+1}\|_F \leq \epsilon$ and
\NoNumber{$\sum_{k=0}^{K-1}\|\mathbf G_k^{i+1}-\mathbf G_k^{i}\|_F$ $\leq \epsilon$.}
\EndFor
\end{algorithmic}
\end{algorithm}

\subsection{$\mathbf L$-minimization using the Anderson-Moore method}
\label{subsec: L}
In this section, we apply the Anderson-Moore method to the $\mathbf L$-minimization step (\ref{eq: L1}).
The Anderson-Moore method is an iterative technique for solving systems of coupled matrix equations efficiently.
We refer the reader to \cite{FMM2013J} for a more detailed discussion of its applications and related references. In what follows, we extend the approach of \cite{FMM2013J} to account for the periodicity of the sensor schedule.

Completing the squares with respect to $\{\mathbf L_k \}$ in the augmented Lagrangian (\ref{eq: ALag1}), the $\mathbf L$-minimization step in (\ref{eq: L1}) can be expressed as \cite{SNEn2011,FMM2013J}
\begin{align}
\begin{array}{ll}
\minimize
&
\displaystyle{\sum_{k=0}^{K-1}} f_k(\{\mathbf L_k\}) + \displaystyle{\sum_{k=0}^{K-1}} \frac{\rho}{2}  || \mathbf L_k - \mathbf U_k^i ||_F^2
\end{array}
\label{eq: L1_obj}
\end{align}
where $\mathbf U_k^i := \mathbf G_k^i - (1/\rho)\boldsymbol \Lambda_k^i$ for $k=0,1,\ldots,K-1$. For notational simplicity, henceforth we will use $\mathbf U_k$ instead of $\mathbf U_k^i$, where $i$ indicates the iteration index.
We bring attention to the fact that, by defining the indicator function $\mathcal{I}$ in (\ref{eq: indicator}) and then splitting the optimization variables in (\ref{eq: obj_1_admm}), we have effectively removed both sparsity penalties and energy constraints from the variables $\{\mathbf L_k\}$ in the $\mathbf L$-minimization problem (\ref{eq: L1_obj}). This is a key advantage of applying ADMM to the sensor scheduling problem.

Recalling the definition of $f_k$, problem (\ref{eq: L1_obj}) can be equivalently written as
\[
\hspace*{-0.03in}
\begin{array}{ll}
\text{minimize}
& \hspace*{-0.08in}
\phi(\{\mathbf L_k\}) :=
\displaystyle{\sum_{k=0}^{K-1}} \tr(\mathbf P_k) + \displaystyle{\sum_{k=0}^{K-1}} \frac{\rho}{2}  || \mathbf L_k - \mathbf U_k ||_F^2
\\[0.15cm]
\text{subject to} & \hspace*{-0.08in}
\left \{
\begin{array}{l}
\hspace*{-0.05in}\text{Lyapunov recursion (\ref{eq: P_lyapunov}) for $k = 0, 1, \ldots, K\!-\!1$,}\\[0.15cm]
\hspace*{-0.05in}\text{periodicity condition (\ref{eq: period}).}
\end{array} \right.
\end{array}
\]

\begin{myprop}
\it{
The necessary conditions for the optimality of a sequence $\{ \mathbf L_k \}$ can be expressed as the set of coupled matrix recursions
\begin{align*}
\mathbf P_{k+1} &= (\mathbf A - \mathbf L_k \mathbf C)  \mathbf P_k (\mathbf A - \mathbf L_k \mathbf C)^T + \mathbf B \mathbf Q \mathbf B^T + \mathbf L_k \mathbf R \mathbf L_k^T  
\\
\mathbf V_k &= (\mathbf A-\mathbf L_k \mathbf C)^T \mathbf V_{k+1} (\mathbf A - \mathbf L_k \mathbf C) + \mathbf I  
\\
\mathbf 0 &= 2 \mathbf V_{k+1}  \mathbf L_k \mathbf R - 2 \mathbf V_{k+1} (\mathbf A - \mathbf L_k \mathbf C) \mathbf P_k \mathbf C^T \!\!+ \rho (\mathbf L_k - \mathbf U_k) 
\end{align*}
for $k =0,\ldots, K-1$, where $\mathbf U_k := \mathbf G_k^i - (1/ \rho) \boldsymbol \Lambda_k^i $ and $\mathbf L_K = \mathbf L_0$, $\mathbf P_K = \mathbf P_0$. The expression on the right of the last equation is the gradient of $\phi$ with respect to $\mathbf L_k$.
}
\end{myprop}
\textbf{Proof:}
See appendix \ref{sec: append1}.  \hfill$\blacksquare$

Due to their coupling, it is a difficult exercise to solve the above set of matrix equations. We thus employ the Anderson-Moore method \cite{FMM2013J,FMM2011_2}, which is an efficient technique for iteratively solving systems of coupled Lyapunov and Sylvester \textit{equations}. We note, however, that the set of matrix equations given in the proposition include (periodic) Lyapunov {\em recursions} rather than (time-independent) Lyapunov equations.
We next apply what can be thought of as a {\em lifting} procedure \cite{chen} to take the periodicity out of these equations and place them in a form appropriate for the application of the Anderson-Moore method.

Let $ \mathbfcal{T}$ denote the following permutation matrix in block-cyclic form \cite{HD2006}
\[
\mathbfcal{ T} := \begin{bmatrix}
      \mathbf{0} &  &  & \mathbf I   \\
     \mathbf  I  & \ddots & &    \\
         & \ddots & \ddots & \\
         & & \mathbf I & \mathbf{0}
     \end{bmatrix}
\]
where $\mathbf I$ is a $N \times N$ identity matrix,
and define
\begin{align*}
\mathbfcal{L} &:= \mathbfcal{T} \mathrm{diag}\{\mathbf L_k \} = \begin{bmatrix}
      \mathbf{0} &  &  & \mathbf L_{K-1}          \\
      \mathbf L_0  &  & &    \\
             & \ddots & \ddots & \\
         & & \mathbf L_{K-2} & \mathbf{0}
     \end{bmatrix},
\\
\mathbfcal{P} &:= \mathrm{diag}\{\mathbf P_k\},
~~
\mathbfcal{V} := \mathrm{diag}\{\mathbf V_k\},
~~
\mathbfcal{U} := \mathcal{T} \mathrm{diag} \{\mathbf U_k \},
\\
\mathbfcal{Q} &:= \mathrm{diag} \{\mathbf Q\},
~~
\mathbfcal{R} := \mathrm{diag} \{\mathbf R \},
~~
\mathbfcal{I} := \mathrm{diag}\{\mathbf I\},
\\
\mathbfcal{A} &:= \mathcal{T} \mathrm{diag}\{\mathbf A \},
~~
\mathbfcal{B} := \mathrm{diag} \{\mathbf B \},
~~
\mathbfcal{C} := \mathrm{diag} \{\mathbf C \}.
\end{align*}
In the sequel, we do not distinguish between the sequence $\{\mathbf L_k \}$ and its cyclic form $\mathcal{L}$, and will alternate between the two representations as needed.
The recursive equations in the statement of Proposition\,$1$ can now be rewritten in the time-independent form
\begin{align}
\mathbfcal{P} &= (\mathbfcal{A} - \mathbfcal{L}\mathbfcal{C}) \mathbfcal{P}  (\mathbfcal{A} - \mathbfcal{L}\mathbfcal{C})^T + \mathbfcal{B}\mathbfcal{Q}\mathbfcal{B}^T + \mathbfcal{L} \mathbfcal{R} \mathbfcal{L}^T 
\label{eq:NC-P'} \\
\mathbfcal{V} &= (\mathbfcal{A} - \mathbfcal{L}\mathbfcal{C})^T \mathbfcal{V} (\mathbfcal{A}-\mathbfcal{L}\mathbfcal{C})+ \mathbfcal{I} 
\label{eq:NC-V'} \\
\mathbf 0 &= 2 \mathbfcal{V} \mathbfcal{L} \mathbfcal{R} - 2 \mathbfcal{V} (\mathbfcal{A} - \mathbfcal{LC})\mathbfcal{PC}^T + \rho(\mathbfcal{L} - \mathbfcal{U}) 
\label{eq:NC-L'}
\end{align}
Furthermore, defining
\[
\nabla \mathbf \Phi := \mathbfcal{T} \mathrm{diag} \{ \nabla_{\!\mathbf L_k} \phi \} = \begin{bmatrix}
      \mathbf{0} &  &  & \nabla_{\!\mathbf L_{K-1}} \phi         \\
     \nabla_{\!\mathbf L_{0}} \phi       & \ddots &  & \\
  & \ddots & \ddots & \\
         & & \nabla_{\!\mathbf L_{K-2}} \phi  & \mathbf{0}
     \end{bmatrix}
\]
it can be shown that
\begin{align}
\nabla \mathbf \Phi = 2 \mathbfcal{VLR} - 2 \mathbfcal{V(A-LC)PC^T} + \rho (\mathbfcal{L-U}),
\label{eq: gradient_Phi}
\end{align}
i.e., the right side of (\ref{eq:NC-L'}) gives the gradient direction for $\mathbfcal{L}$, or equivalently the gradient direction for each $\mathbf L_k$, $k=0,1,\ldots,K-1$.

We briefly describe the implementation of the Anderson-Moore method as follows.
For each iteration of this method, we first keep the value of $\mathbfcal{L}$ fixed and solve (\ref{eq:NC-P'}) and (\ref{eq:NC-V'}) for $\mathbfcal{P}$ and $\mathbfcal V$, then keep $\mathbfcal{P}$ and $\mathbfcal V$ fixed and solve (\ref{eq:NC-L'}) for a new value $\mathbfcal L_{new}$ of $\mathbfcal L$. Proposition\,\ref{prop: direction} shows that the difference $\tilde {\mathbfcal L} := \mathbfcal L_{new} - \mathbfcal L$ between the values of $\mathbfcal L$ over two consecutive iterations constitutes a descent direction for $\phi(\{ \mathbf L_k \})$; see \cite{FMM2013J,FMM2011_2} for related results. We employ a line search \cite{Boyd2004_bk} to determine the step-size $s$ in $\mathbfcal L + s \tilde {\mathbfcal L}$ in order to accelerate the convergence to a stationary point of $\phi$. 
We also assume that there always exists an $\mathbfcal L$ that satisfies the measurement frequency constraint and for which the spectrum of $\mathbfcal A - \mathbfcal L \mathbfcal C$ is contained inside the open unit disk; we elaborate on this condition in Section \ref{initialize.sec}. These assumptions guarantee the existence of unique positive definite solutions $\mathbfcal P$ and $\mathbfcal V$ to Equations (\ref{eq:NC-P'}) and (\ref{eq:NC-V'}) \cite{TNM1982}. 

\begin{myprop}
\label{prop: direction}
The difference $\tilde{\mathbfcal L}: = \mathbfcal L_{new} - \mathbfcal L$ constitutes a descent direction for $\phi (\{ \mathbf L_k\})$,
\begin{align}
 \langle {\nabla \mathbf \Phi, \tilde {\mathbfcal{L}}}  \rangle < 0,
\end{align}
where $ \langle {\nabla \mathbf \Phi, \tilde {\mathbfcal{L}}}  \rangle : = \mathrm{\tr}(\nabla \mathbf \Phi^T \tilde {\mathbfcal{L}} ) = \sum_{k=0}^{K-1}\mathrm{tr}(\nabla_{\!\mathbf L_k} \phi^T \mathbf L_k)$. Moreover, $\langle {\nabla \mathbf \Phi(\mathbfcal L), \tilde{\mathbfcal L}} \rangle = 0$ if and only if $\mathbfcal L$ is a stationary point of $\mathbf \Phi$, i.e., $\nabla \mathbf \Phi (\mathbfcal L) = \mathbf 0$.
\end{myprop}
\textbf{Proof:}
The proof is similar to \cite[Prop.\,$1$]{FMM2011_TechRep} and omitted for brevity. \hfill $\blacksquare$ 

We summarize the Anderson-Moore method for solving the $\mathbf L$-minimization step in Algorithm\,2. This algorithm calls on the Armijo rule \cite{DP_book}, given in Algorithm\,3, to update $\mathbfcal L$.

\begin{algorithm}
\caption{$L$-minimization step (\ref{eq: L1}), in the $i$th iteration of ADMM, using Anderson-Moore}
\begin{algorithmic}[1]
\State If $i = 0$, choose $\mathbfcal L^0$ from (\ref{eq: iniL}). If $i \geq 1$, set $\mathbfcal L^0$ equal to solution of (\ref{eq: L1}) from previous ADMM iteration.
\For{$t=0,1,\ldots$}
\State Set $\mathbfcal L = \mathbfcal L^t$ and solve (\ref{eq:NC-P'}), (\ref{eq:NC-V'}) to find $\mathbfcal P^t$, $\mathbfcal V^t$.
\State Set $\mathbfcal V = \mathbfcal V^t$, $\mathbfcal P = \mathbfcal P^t$ and solve (\ref{eq:NC-L'}) to find $\bar {\mathbfcal L^t} $.
\State Compute $\tilde {\mathbfcal L^t} = \bar {\mathbfcal L^t} - \mathbfcal L^t$ and update
$\mathbfcal L^{t+1} = \mathbfcal L^t+ s^t \tilde {\mathbfcal L^t}$,
\hspace*{0.4cm} where $s^t$ given by Armijo rule (see Algorithm\,3).
\State \textbf{until} $\| \nabla \mathbf \Phi(\mathbfcal L^t) \| < \epsilon$.
\EndFor
\end{algorithmic}
\end{algorithm}
\begin{algorithm}
\caption{Armijo rule for choosing step-size $s^t$}
\begin{algorithmic}[1]
\State Set $s^t = 1$ and choose $\alpha, \beta \in (0,1)$.
\Repeat
\State $s^t = \beta s^t$,
\Until $\phi(\mathbfcal L^t + s^t \tilde{\mathbfcal L^t}) < \phi(\mathbfcal L^t) + \alpha \, s^t \tr\!\big(\nabla \mathbf \Phi(\mathbfcal L^t)^T \tilde{\mathbfcal L^t}\big)$.
\end{algorithmic}
\end{algorithm}

\subsection{$\mathbf G$-minimization}

In this section, we consider the $\mathbf G$-minimization step (\ref{eq: G1}) and demonstrate that it can be solved analytically. In what follows, we extend the approach of \cite{FMM2013J} to account for  the periodicity and energy constraints in the sensor schedule.

Completing the squares with respect to $\{\mathbf G_k \}$ in the augmented Lagrangian (\ref{eq: ALag1}), the $\mathbf G$-minimization step in (\ref{eq: G1}) can be expressed as \cite{SNEn2011,FMM2013J}
\[
\begin{array}{ll}
\minimize
&
\gamma \displaystyle{\sum_{k=0}^{K-1}} g(\mathbf G_k) + \frac{\rho}{2} \displaystyle{\sum_{k=0}^{K-1}} ||\mathbf G_k -\mathbf S_k^i ||_F^2
\\[0.15cm]
\text{subject to}
& \displaystyle{\sum_{k=0}^{K-1}} \card\! \big(  \|\mathbf G_{k,m} \|_2  \big) \leq \eta,
~~~
m=1,2,\ldots, M,
\end{array}
\]
where $\mathbf S_k^i := \mathbf L_k^{i+1} + (1/\rho)\boldsymbol \Lambda_k^i$ for $k=0,1,\ldots,K-1$. 
For notational simplicity, henceforth we will use $\mathbf S_k$ instead of $\mathbf S_k^i$, where $i$ indicates the iteration index.
Recalling the definition of $g$ from (\ref{eq: card_g0}), and replacing $||\mathbf G_k -\mathbf S_k ||_F^2$ with $\sum_{m=1}^M \|\mathbf  G_{k,m} - \mathbf S_{k,m} \|_2^2 $ yields the equivalent optimization problem
\[
\begin{array}{ll}
\text{minimize}
&
\psi(\{\mathbf G_k\}) :=
\displaystyle{\sum_{m=1}^{M}} \Big( \displaystyle{\sum_{k=0}^{K-1}} \gamma \card\! \big(  \| \mathbf G_{k,m}\|_2  \big)
\\[0.15cm]
& \hspace{1in}
+ \displaystyle{\sum_{k=0}^{K-1}} \frac{\rho}{2}  \| \mathbf G_{k,m} - \mathbf S_{k,m} \|_2^2 \Big)
\\[0.15cm]
\text{subject to}
& \displaystyle{\sum_{k=0}^{K-1}} \card\! \big(  \|\mathbf G_{k,m} \|_2  \big) \leq \eta,
~~~
m=1,2,\ldots, M,
\end{array}
\]
where we have exploited the {\em column-wise separability} of $g(\cdot)$ and that of the Frobenius norm.

We form the matrix $\mathbfcal{G}_m$ by picking out the $m$th column from each of the matrices in the set $\{\mathbf G_k \}$ and stacking them,
$
\mathbfcal{G}_m
:=
\begin{bmatrix}
      \mathbf G_{0,m} & \mathbf G_{1,m} & \cdots & \mathbf G_{K-1,m}
\end{bmatrix}.
$
Then the $\mathbf G$-minimization problem decomposes into the subproblems
\begin{equation}
\!\!\!\!\!
\begin{array}{ll}
\text{minimize}
&
\psi_m(\mathbfcal G_m)
:=
\displaystyle{\sum_{k=0}^{K-1}} \gamma \card\! \big(  \|\mathbf G_{k,m}\|_2 \big)
\\[0.15cm]
& \hspace{0.75in}
+ \displaystyle{\sum_{k=0}^{K-1}} \frac{\rho}{2} \| \mathbf G_{k,m}- \mathbf S_{k,m} \| _2^2
\\[0.15cm]
\text{subject to}
& \displaystyle{\sum_{k=0}^{K-1}} \card\! \big(  \| \mathbf G_{k,m} \|_2  \big) \leq \eta,
\label{eq: subG1_ineq}
\end{array}
\end{equation}
which can be solved separately for $m=1,2,\ldots, M$.

To solve problem (\ref{eq: subG1_ineq}) we rewrite the feasible set $F$ of (\ref{eq: subG1_ineq}),
$
F = \big \{ \mathbfcal G_m: \sum_{k=0}^{K-1} \card\! \big(  \|\mathbf G_{k,m}\|_2  \big) \leq \eta \big \},
$
as the union $F = F_{0} \cup F_{1} \cup \cdots \cup F_{\eta}$ of the smaller sets $F_{q}$, $q=0,\ldots,\eta$,
\[
F_{q} = \big \{ \mathbfcal G_m : \sum_{k=0}^{K-1} \card\! \big(  \| \mathbf G_{k,m}\|_2  \big) = q \big \}.
\]
Let $\mathbfcal G_m^q$ denote a solution of
\begin{equation}
\begin{array}{ll}
\text{minimize}
&
\psi_m(\mathbfcal G_m)
\\[0.15cm]
\text{subject to}
& \mathbfcal G_m \in F_{q}.
\label{eq: subG2_ineq}
\end{array}
\end{equation}
Then a minimizer of (\ref{eq: subG1_ineq}) can be obtained by comparing $\psi_m(\mathbfcal G_m^q)$ for $q=0,\ldots,\eta$ and choosing the one with the least value.
The above procedure, together with finding the solution of (\ref{eq: subG2_ineq}), is made precise by the following proposition.
\begin{myprop}
\label{prop_ineq}
The solution of (\ref{eq: subG1_ineq}) is obtained by solving the sequence of minimization problems (\ref{eq: subG2_ineq}) for $q = 0,1,\ldots, \mathrm{min}\{\eta,\kappa\}$,
$\kappa = \sum_{k=0}^{K-1} \card\! \big(  \|\mathbf S_{k,m}\|_2  \big)$.
Furthermore, the solution of (\ref{eq: subG2_ineq}) is given by
\[
\mathbf G_{k,m} = \left\{
  \begin{array}{l l}
    \mathbf S_{k,m} & \quad || \mathbf S_{k,m}||_2 \geq || [\mathbfcal S_m]_{q}||_2 \text{ and } q \neq 0,    \\[0.15cm]
 \mathbf{0} & \quad  \text{otherwise}, 
  \end{array} \right.
\]
for $k = 0,1, \cdots,K-1$, where $\mathbf S_k := \mathbf L_k^{i+1} + (1/\rho)\boldsymbol \Lambda_k^i$,
$\mathbfcal S_m:= [ \mathbf S_{0,m}, \cdots, \mathbf S_{K-1,m} ]$, $[\mathbfcal S_m]_{q}$ denotes the $q$th largest column of $\mathbfcal S_m$ in the $2$-norm sense, and $\mathbf G_{k,m}$, $\mathbf S_{k,m}$ denote the $m$th columns of $\mathbf G_k$, $\mathbf S_k$, respectively.
\end{myprop}
\textbf{Proof:} See Appendix \ref{sec: append3}. \hfill$\blacksquare$

We note that problem (\ref{eq: subG1_ineq}) can be solved via a sequence of equality constrained problems (\ref{eq: subG2_ineq}) whose analytical solution is determined by Proposition\,\ref{prop_ineq}.
However, instead of solving $\mathrm{min}\{\eta,\kappa\} + 1$ equality constrained problems, it is shown in Proposition\,\ref{prop_G2} that the solution of the $G$-minimization problem (\ref{eq: subG1_ineq}) is determined by
the magnitude of the sparsity-promoting parameter $\gamma$.

\begin{myprop}
\label{prop_G2}
The solution $\mathbfcal{G}_m$ of (\ref{eq: subG1_ineq}) is determined by 
solving one subproblem (\ref{eq: subG2_ineq}) based on the value of $\gamma$,
\begin{align}
\mathbfcal G_m = \left \{
\begin{array}{lrcccl}
\mathbfcal G_m^0 \hspace*{-0.2in}
& \frac{\rho}{2} \| [\mathbfcal S_m]_{1} \|_2^2 
\!\!&\!\!<\!\!&\!\! 
\gamma & & \\
\mathbfcal G_m^1 \hspace*{-0.2in}
& ~~ \frac{\rho}{2} \| [\mathbfcal S_m]_{2} \|_2^2 
\!\!&\!\!<\!\!&\!\! 
\gamma 
\!\!&\!\!\leq\!\!&\!\! 
\frac{\rho}{2} \| [\mathbfcal S_m]_{1} \|_2^2 \\
\vdots & & & \vdots & & \\
\mathbfcal G_m^{\mathrm{min}\{\eta, \kappa \}} \hspace*{-0.2in}
& & & \gamma 
\!\!&\!\!\leq\!\!&\!\! 
\frac{\rho}{2} \| [\mathbfcal S_m]_{\mathrm{min}\{ \eta, \kappa\}} \|_2^2
\end{array}
\right. 
\end{align}
where $\mathbfcal G_m^q$ denotes a solution of (\ref{eq: subG2_ineq}) with $q = 0,1,\ldots, \mathrm{min}\{\eta, \kappa \} $, and $\kappa$ and $[\mathbfcal S_m]_{q}$ are defined as in Proposition\,\ref{prop_ineq}.
\end{myprop}
\textbf{Proof:}
See Appendix \ref{sec: append4}. \hfill$\blacksquare$

It is clear from Proposition\,\ref{prop_G2} that the parameter $\gamma$ governs the column-sparsity of $\mathbfcal G_m$. For example, $\mathbfcal G_m$ becomes the zero matrix as $\gamma \to \infty$, which corresponds to the scenario in which all sensors are always inactive.

To reiterate, in order to solve the $G$-minimization problem (\ref{eq: G1}), we first decompose it into the $M$ subproblems (\ref{eq: subG1_ineq}) with separate optimization variables 
$\{\mathbfcal G_m \}_{m = 1, \ldots, M}$.
Each inequality constrained subproblem (\ref{eq: subG1_ineq}) is then solved via Proposition\,\ref{prop_G2}, in which the solution of the equality constrained problem (\ref{eq: subG2_ineq}) is determined by Proposition\,\ref{prop_ineq}. We summarize this procedure in Algorithm\,4.

\begin{algorithm}
\caption{$\mathbf G$-minimization step (\ref{eq: G1})}
\begin{algorithmic}[1]
\State Given $\eta$ and $\mathbf S_k = \mathbf L_k^{i+1} + 1/\rho \boldsymbol \Lambda_k^i$,
set 
\NoNumber{$\kappa = \sum_{k=0}^{K-1} \card\big(\| \mathbf S_{k,m}\|_2  \big)$.}
\For{$m=1,\ldots,M$}
\State Set $\mathbfcal S_m = [ \mathbf S_{0,m}, \cdots,\mathbf S_{K-1,m} ]$.
\State Solve (\ref{eq: subG1_ineq}) using Prop.\,\ref{prop_G2} to obtain $\mathbfcal G_m = \mathbfcal G_m^q$, 
where 
\hspace*{0.15in}
$\mathbfcal G_m^q$ is determined
from Prop.\,\ref{prop_ineq}.  
\EndFor
\State Use $\{\mathbfcal G_m\}_{m = 1, \ldots, M}$ to construct $\{\mathbf G_k\}_{k = 0, 1,\ldots, K-1}$.
\end{algorithmic}
\end{algorithm}


\subsection{Convergence \& Initialization of ADMM-based periodic sensor scheduling}
\label{initialize.sec}

The solution of ADMM for a nonconvex problem generally yields a locally optimal point, and in general depends on the parameter $\rho$ and the initial values of $\{ L_k \}$ and $\{ G_k \}$ \cite{SNEn2011}. In fact for a nonconvex problem, such as the one considered here, even the convergence of ADMM is not guaranteed\cite{SNEn2011}. Our numerical experiments and those in other works such as \cite{FMM2013J} demonstrate that ADMM indeed works well when the value of $\rho$ is chosen to be large. However, \textit{very} large values of $\rho$ make the Frobenius norm dominate the augmented Lagrangian (\ref{eq: ALag1}) and thus lead to less emphasis on 
minimizing the estimation error.
In order to select an appropriate value of $\rho$, certain extensions (e.g., varying penalty parameter) of the classical ADMM algorithm have been explored. The reader is referred to \cite[Sec.\,$3$]{SNEn2011}.

To initialize the estimator gain $\{ \mathbf L_k \}$, we start with a feasible initializing sensor schedule. Such a schedule can be expressed in terms of the observation matrices over one period, namely, $\mathbf C(k)= [\zeta_{k,1} \mathbf C_1, \ldots, \zeta_{k,M} \mathbf C_M]^T$ for $k = 0,1,\ldots,K-1$, where the binary variable $\zeta_{k,m}$ indicates whether or not the $m$th sensor is active at time $k$. Note that the periodic sensor schedule $\{\mathbf C(k)\}$ uniquely determines the limit cycle of the periodic error covariance matrix \cite{WMJ2010}. 
We express the periodic sensor schedule $\{\mathbf C(k) \}$ in cyclic form 
\begin{align*}
\mathbfcal{C}^0 \hspace*{-0.02in}:= \hspace*{-0.02in}\mathbfcal{T} \; \mathrm{diag}\{ \mathbf C(k) \}\hspace*{-0.02in} =
\hspace*{-0.04in}
 \begin{bmatrix}
      \mathbf{0} &  &  & \mathbf C(K-1)          \\
     \mathbf C(0) &  & &    \\
             & \ddots & \ddots & \\
         & & \mathbf C(K-2) & \mathbf{0}
     \end{bmatrix},
\end{align*}
and solve the following algebraic Riccati equation for the cyclic form of $\{\mathbf P_k \}$
\begin{align}
\mathbfcal P = \mathbfcal Q + \mathbfcal A \mathbfcal P \mathbfcal A^T - \mathbfcal A \mathbfcal P \mathbfcal{C}^{0T} (\mathbfcal{C}^0 \mathbfcal P \mathbfcal{C}^{0T} + \mathbfcal R)^{-1} \mathbfcal{C}^0 \mathbfcal P \mathbfcal A^{-1},
\label{eq: Riccati}
\end{align}
where $\mathbfcal P$, $\mathbfcal Q$, $\mathbfcal A$ and $\mathbfcal R$ have the same definitions as in Sec. \ref{subsec: L}. 
The Riccati equation (\ref{eq: Riccati}) gives the optimal periodic estimator gain corresponding to a discrete-time system with {\em given} periodic observation matrices $\{ \mathbf C(k) \}$.
Once the solution of (\ref{eq: Riccati}) is found, the corresponding estimator gain in cyclic form is given by \cite{Kalman}
\begin{align}
{\mathbfcal L}^0 = \mathbfcal A \mathbfcal P \mathbfcal{C}^{0T} (\mathbfcal{C}^0 \mathbfcal P \mathbfcal{C}^{0T} + \mathbfcal R)^{-1} {\mathbfcal T}^0,
\label{eq: iniL}
\end{align}
where ${\mathbfcal T}^0$ has the same block-cyclic form of $\mathbfcal T$ but is instead formed using $M \!\times\! M$ identity matrices.

It is not difficult to show that the matrix $\mathbfcal{L}^0$ in (\ref{eq: iniL}) has the same sparsity pattern as $\mathbfcal{C}^0$. Thus, the sequence $\{ \mathbf L_k^0 \}$ obtained from $\mathbfcal{L}^0$ respects the energy constraints and can be used to initialize ADMM. Furthermore, we assume that $({\mathbfcal C}^0,\mathbfcal A)$ is observable, which guarantees that the spectrum of $\mathbfcal A - {\mathbfcal L}^0 {\mathbfcal C}^0$ is contained inside the open unit disk and thus the initializing estimator gains $\{\mathbf L_k^0\}$ will be stabilizing. 
Finally, for simplicity $\{\mathbf G_k \}$ is initialized to $\mathbf G_k = \mathbf 0$, $k = 0,1,\ldots,K-1$.

\subsection{Complexity analysis}

It has been shown that ADMM typically takes a few tens of iterations to converge
with modest accuracy for many applications \cite{FMM2011,FMM2013J,EMP2012,FMM2011_2,SNEn2011}. 
The computational complexity of each iteration of ADMM is dominated by the $\mathbf L$-minimization step, since the analytical solution of the $\mathbf G$-minimization step can be directly obtained and the dual update is calculated by matrix addition. For the $\mathbf L$-minimization subproblem, the descent Anderson-Moore method requires the solutions of two Lyapunov equations (\ref{eq:NC-P'})-(\ref{eq:NC-V'}) and one Sylvester equation (\ref{eq:NC-L'}) at each iteration. 
To solve them, the Bartels-Stewart method \cite{RG72} 
yields the complexity 
$O(K^3 N^3 + K^3M^3 + K^3 M N^2 + K^3 N M^2)$, 
where $K$ is the length of the period, $M$ is the number of sensors and $N$ is the dimension of the state vector. 
We also note that the convergence of the Anderson-Moore method is guaranteed by Prop.\,2, and it typically requires a small number of iterations because of the implementation of the Armijo rule. 

For additional perspective, we compare the computational complexity of our proposed methodology to a periodic sensor scheduling problem that is solved by semidefinite programming (SDP), for example as done in \cite{TM2011}. The complexity of SDP
is approximated by $O(a^2 b^{2.5} + a b^{3.5})$ \cite{Interior_bk}, where $a$ and $b$ denote the number of optimization variables and the size of the semidefinite matrix, respectively. For the linear matrix inequality (LMI) problem proposed in \cite{TM2011}, the computation complexity
is determined by $a = N(N+1)/2+M$ and $b = (K+1)N + M$. 
Thus, problems involving large-scale dynamical system with many state variables, result in large SDPs with computation complexity $O(N^{6.5})$. It can be seen that our approach reduces the computational complexity by a factor of $N^{3.5}$ compared to the LMI-based method of \cite{TM2011}.


\section{Example: Field Estimation of a Spatially Extended System}
\label{sec: sim}

In order to demonstrate the effectiveness of our proposed periodic sensor scheduling algorithm, we consider the example of field monitoring. In this problem, sensors are deployed on a 
rectangular region to estimate the state of a diffusion process described by the partial differential equation \cite{HJB2009,YRB2011} 
\begin{subequations}
\begin{align}
& \begin{array}{lll}
& \displaystyle \frac{\partial \xi(\mathbf s, t)}{\partial t} 
= \nabla^2 \xi(\mathbf s, t) & 
\end{array} \\
& \begin{array}{lll}
\hspace*{-1.in}\text{with Dirichlet boundary conditions} &  & 
\end{array} \nonumber \\
& \begin{array}{lll}
& \displaystyle \xi(\mathbf s, \cdot\, ) = 0 \quad \mathbf s \in \partial \mathcal D &
\end{array}
\end{align}
\label{eq: heat_eq}
\end{subequations}
\hspace*{-0.11in} where $\xi(\mathbf s, t)$ denotes the field (or state) value at location $\mathbf s$ and time $t$, $\nabla^2$ denotes the Laplace operator, and $\partial \mathcal D$ denotes the boundary of a {rectangular} region of interest $\mathcal D$.

\begin{figure}[htb]
\centering
\includegraphics[width=.5\textwidth,height=!]{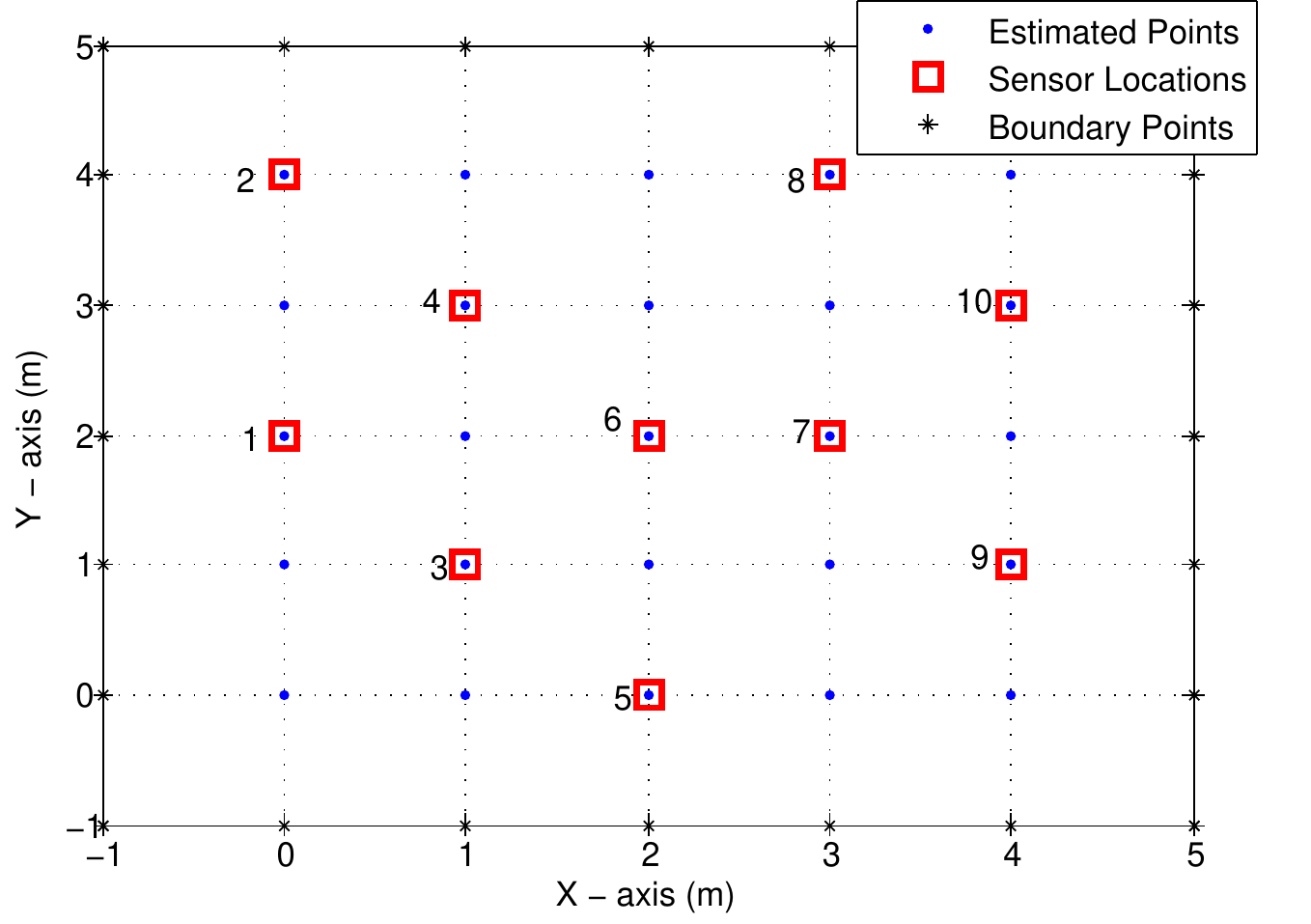}
\caption{$M = 10$ sensors deployed in {a $6 \times 6$ region}.} 
\label{fig:sens_field}
\end{figure}

We consider a spatially-discretized approximation of (\ref{eq: heat_eq}) and our aim is to estimate the state over the entire discrete lattice using a small number of sensors; see Fig.\,\ref{fig:sens_field} for an example.
With an abuse of notation, a simple discrete approximation of (\ref{eq: heat_eq}) can be generated by setting \cite{YRB2011}
\begin{align}
\nabla^2 \xi(\mathbf s, t) \big|_{s=(i,j)} 
& \approx
\frac{\xi(i+1, j, t)-2\xi(i,j,t)+\xi(i-1,j,t)}{h^2} \nonumber \\
& ~ 
+ \frac{\xi(i,j+1,t)-2\xi(i,j,t)+\xi(i,j-1,t)}{h^2},
\label{eq: partial}
\end{align} 
for $i = 0, 1, \ldots, \ell_h$ and $j = 0, 1, \ldots, \ell_v$, where
$\ell_h + 2$ and $\ell_v+2$ are the width and length of a rectangular region, respectively; for example, $\ell_h = \ell_v = 4$ in Fig.\,\ref{fig:sens_field}. 
In (\ref{eq: partial}),
  $h$ denotes the physical distance between the lattice points, and $\xi(-1,j, t) = \xi(\ell_h+1,j, t) = \xi(i,-1, t) = \xi(i,\ell_v+1, t) = 0$ for all indices $i, j$ and time $t$.

From (\ref{eq: heat_eq}) and (\ref{eq: partial}), we can obtain the evolution equations 
$\frac{d}{d t} \mathbf x(t) =  \mathbf A_\Delta {{x}}(t)$, 
where $\mathbf x(t) \in \bbR^N$, $N = (\ell_h+1) \!\times\! (\ell_v+1)$, denotes the state vector 
$
\mathbf x(t) = [\xi(0,0,t), \xi(0,1,t),\ldots,\xi(\ell_h,\ell_v,t)]^T,
$
and $\mathbf A_\Delta$ can be directly computed from (\ref{eq: partial}). Finally, applying a discretization in time and introducing process noise (i.e., a spatio-temporal random field) into the evolution yields
\begin{align}
{\mathbf x}_{k+1} = \mathbf A {\mathbf x}_k + {\mathbf w}_k. \nonumber
\end{align}
Here, ${\mathbf x}_k$ is the state vector, $\mathbf w_k$ is a white Gaussian process with zero mean and covariance matrix $\mathbf Q$, $\mathbf A$ is the system transition matrix  $\mathbf A=e^{\mathbf A_\Delta T}$, and $T$ is the temporal sampling interval.


We assume that $M$ sensors, $M < N$, are deployed and make measurements of the state according to
\begin{align}
{\mathbf y}_k=\mathbf C{\mathbf x}_k + {\mathbf v}_k, \nonumber
\end{align} 
where $\mathbf y_k \in \bbR^M$ is the measurement vector, $\mathbf v_k$ denotes the measurement noise which is a white Gaussian process with zero mean and covariance matrix $\mathbf R$, and
$\mathbf C$ is the $M \!\times\! N$ observation matrix. For example, the case where the $m$th row of $\mathbf C$ contains only one nonzero entry equal to 1 corresponds to the scenario in which the $m$th entry of $\mathbf y_k$ represents measurements of the field at the location of the $m$th sensor.

We consider an instance in which $M = 10$ sensors are deployed to monitor $N = 25 $ field points shown in Fig.\,\ref{fig:sens_field}. We assume that each sensor can be selected at most $\eta$ times, $\eta \in \{1,\ldots,10 \}$, during any period of length $K = 10$. Furthermore, we select $T = 0.5$,  $\mathbf Q = 0.25 \, \mathbf I$, and $\mathbf R =  \mathbf I$. The ADMM stopping tolerance is $\epsilon = 10^{-3}$. In our computations, ADMM converges for $\rho \geq 10$ and the required number of ADMM iterations is approximately $20$.


\begin{figure}[htb]
\centering
\includegraphics[width=.5\textwidth,height=!]{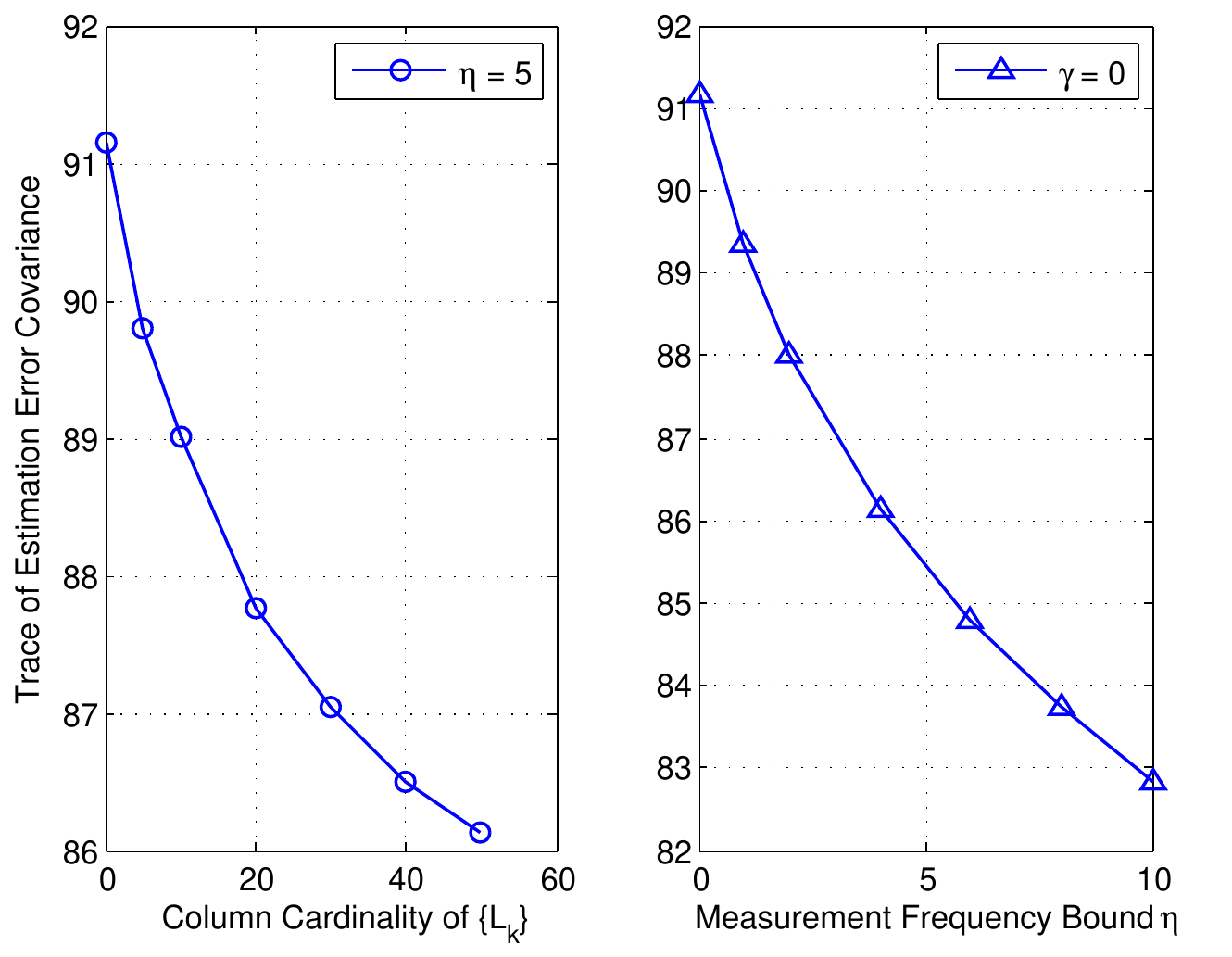} 
\caption{\footnotesize{Estimation performance obtained from our approach.
	Left plot: Tradeoff between estimation performance and total number of sensors (in terms of column-cardinality of $\{ L_k \}$) for a fixed $\eta = 5$;
	Right plot: Estimation performance as a function of measurement frequency bound $\eta$.}
}
\label{fig: Perf_Pareto}
\end{figure}

In Fig.\,\ref{fig: Perf_Pareto}, for our approach we present the estimation performance, namely the cumulative traces of error covariance matrices over one period, respectively as a function of the cumulative column-cardinality of $\{L_k\}$ and the measurement frequency bound $\eta$. In the left plot, we fix $\eta = 5$ and vary $\gamma$, which results in changes in the column-cardinality of $\{L_k\}$ and renders the trade-off curve between the conflicting objectives of good estimation performance and minimal sensor usage. Numerical results demonstrate that as the column-cardinality of $\{ L_k\}$ increases and more sensors are activated, the estimation performance improves.
In the right plot, 
we observe that the estimation performance is improved by increasing $\eta$. This is not surprising, as a larger value of $\eta$ allows the (most informative) sensors to be active more frequently.

\begin{figure}[htb]
\centering
\includegraphics[width=.5\textwidth,height=!]{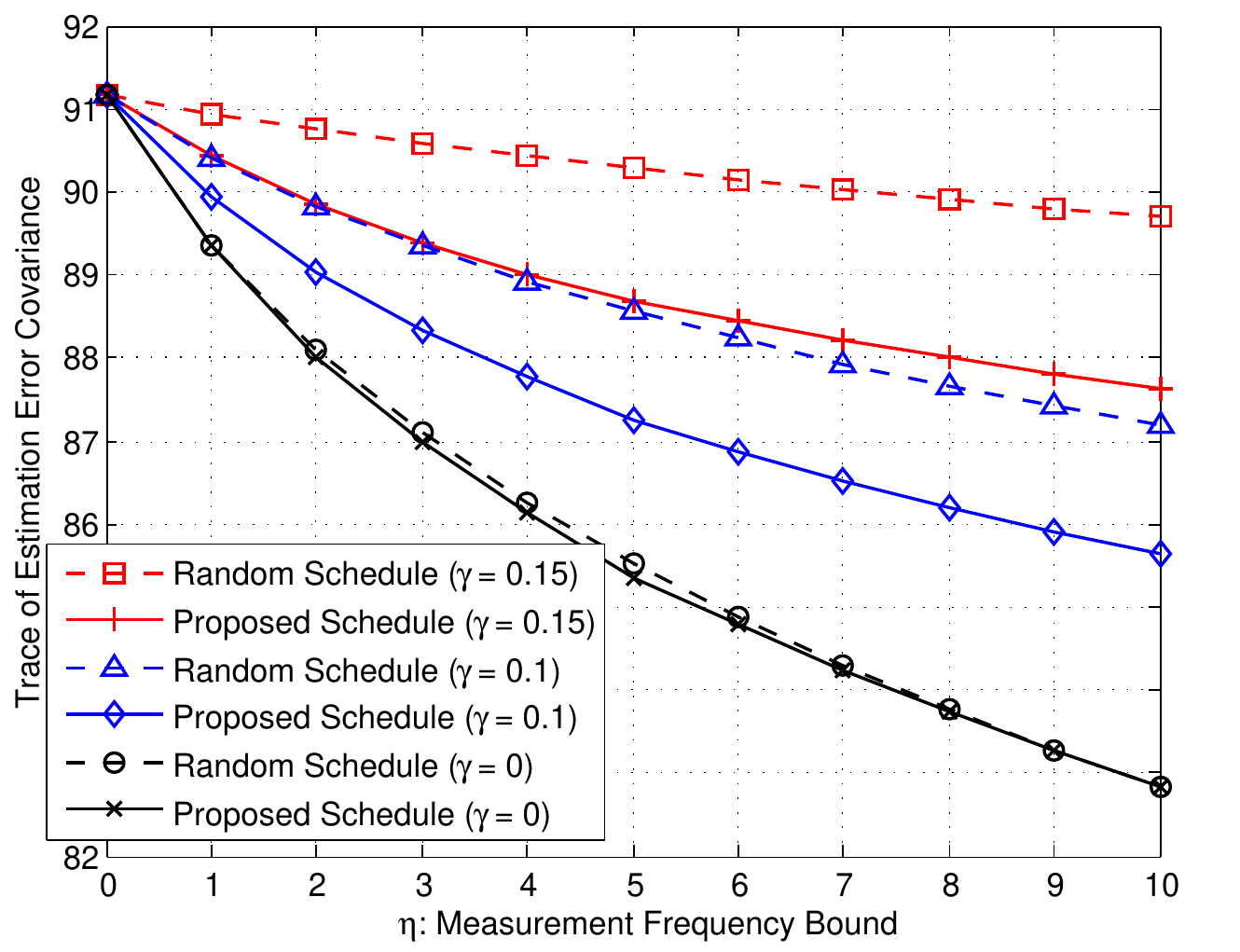} 
\caption{\footnotesize{
Performance comparison of random schedules versus our proposed schedule.
}}
\label{fig: perfcomp_S10K10}
\end{figure}

Next, we compare the estimation performance of our approach to that of random scheduling, where the latter method refers to randomly selected sensor schedules that satisfy the measurement frequency constraint and have the same total number of active sensors over one period as the schedule obtained from our approach. The performance of the random strategy is taken to be the average of the traces of error covariance matrices over $500$ simulation trials. 
In Fig.\,\ref{fig: perfcomp_S10K10}, the estimation performance is presented as a function of the measurement frequency bound $\eta$ for three different values of the sparsity-promoting parameter $\gamma = 0, 0.1, 0.15$. Numerical results show that our approach significantly outperforms the random strategy for $\gamma = 0.1, 0.15$, as the former approach takes into account sensor activations over both time and space. For $\gamma = 0$ there is no penalty on sensor activations, and to achieve the best estimation performance every sensor is active $\eta$ times per period (i.e., all sensors attain their measurement frequency bound). As a consequence, the performance gap between our approach and that of the random strategy is not as large for $\gamma = 0$ as it is for $\gamma > 0$. 
{In our numerical experiments for smaller versions of this example, where exhaustive searches are feasible, we observed that our proposed method yields sensor schedules that are identical or close in performance to the globally optimal schedule found via an exhaustive search.}

\begin{figure*}[htb]
\centerline{ \begin{tabular}{ccc}
\includegraphics[width=.3\textwidth,height=!]{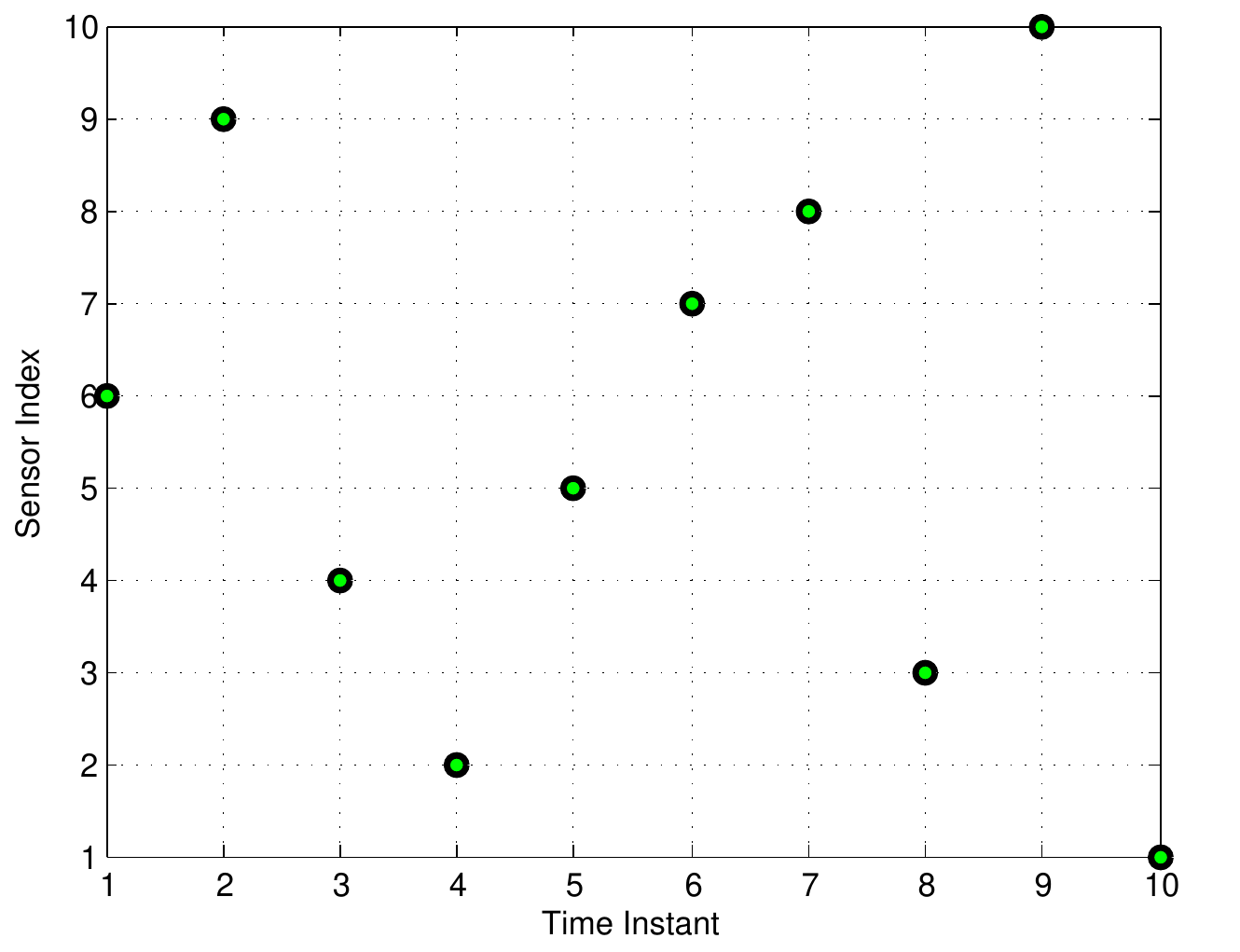}  &
\includegraphics[width=.3\textwidth,height=!]{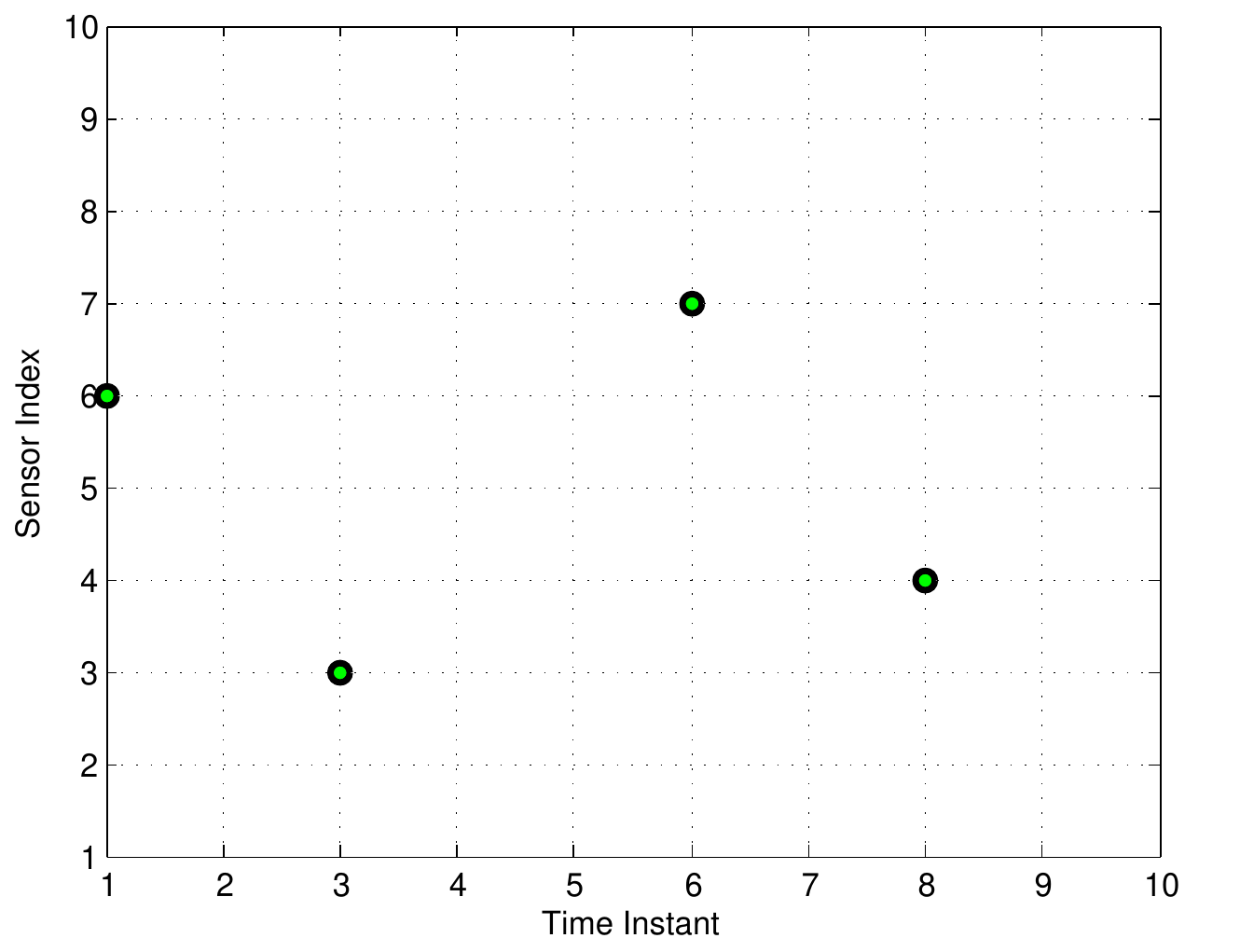}  &
\includegraphics[width=.3\textwidth,height=!]{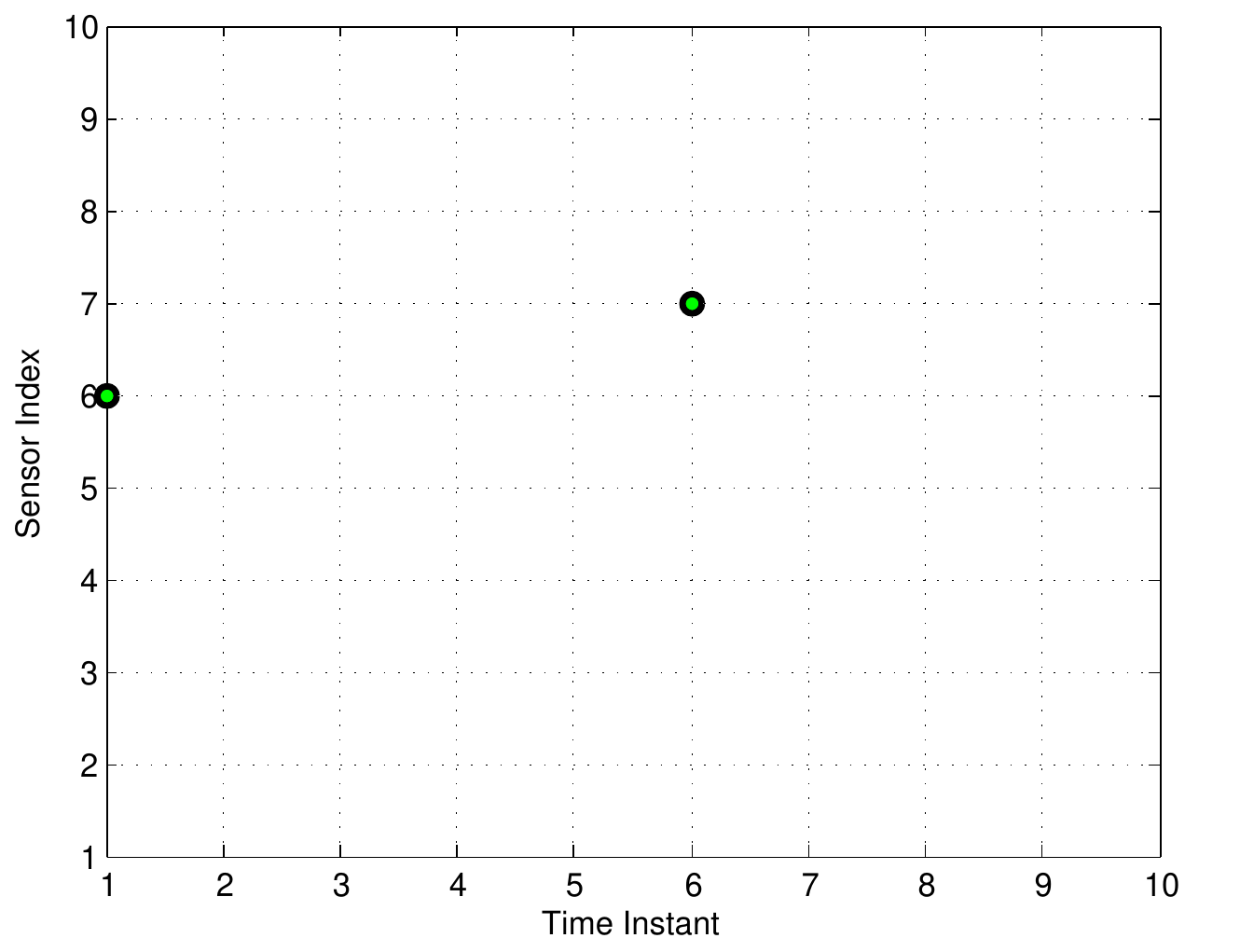} \\
(I-a)&(I-b)&(I-c) \\
\includegraphics[width=.3\textwidth,height=!]{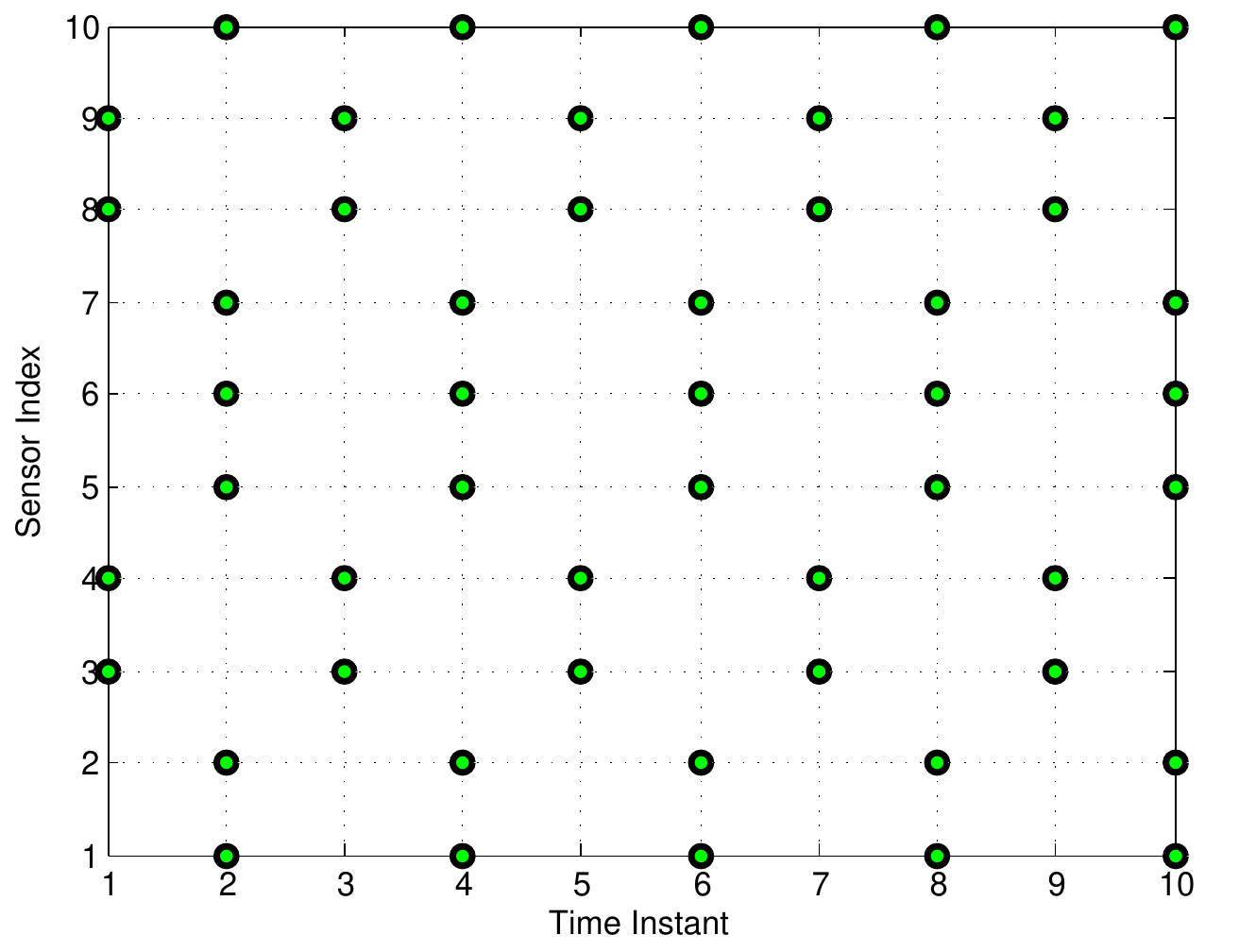} &
\includegraphics[width=.3\textwidth,height=!]{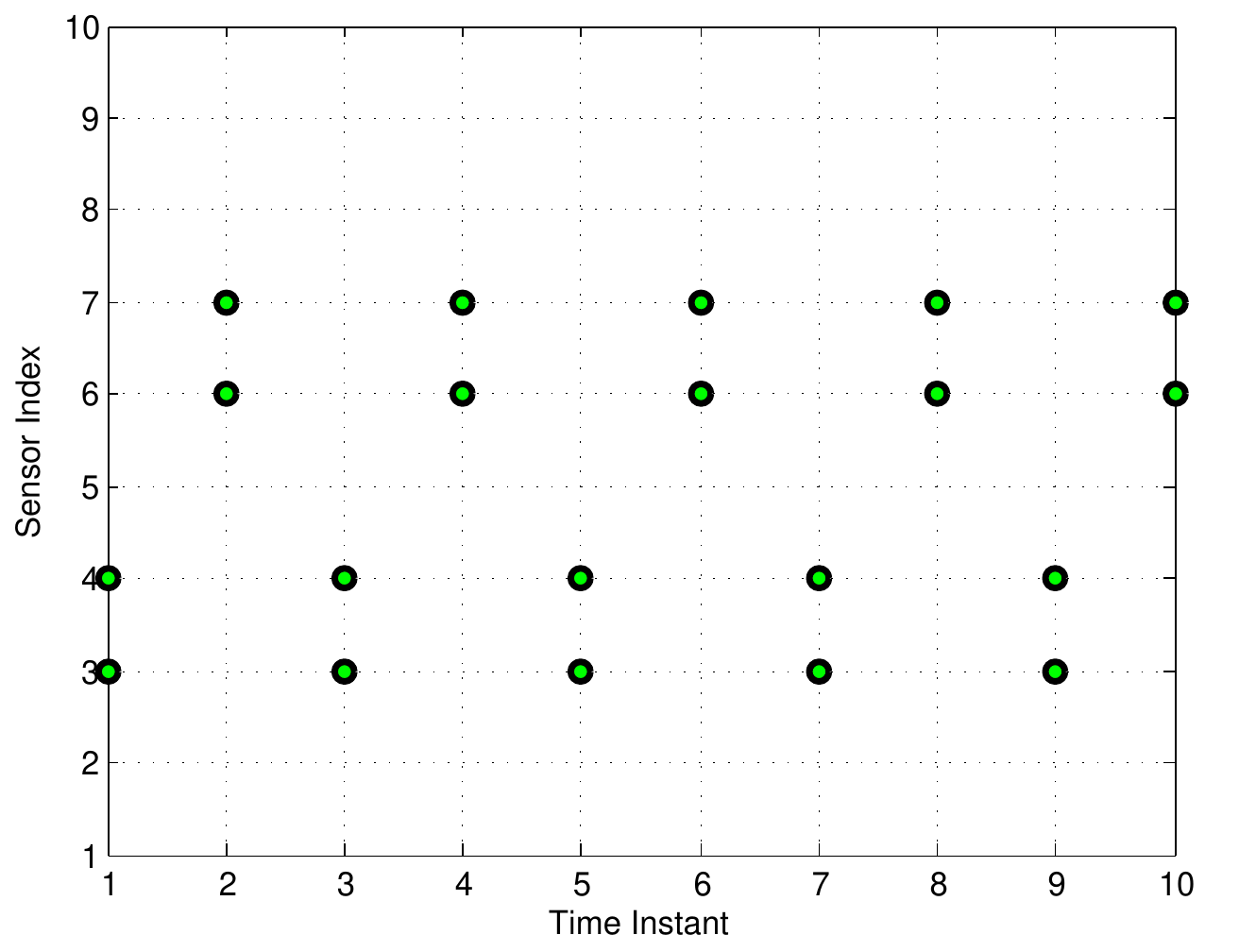} &
\includegraphics[width=.3\textwidth,height=!]{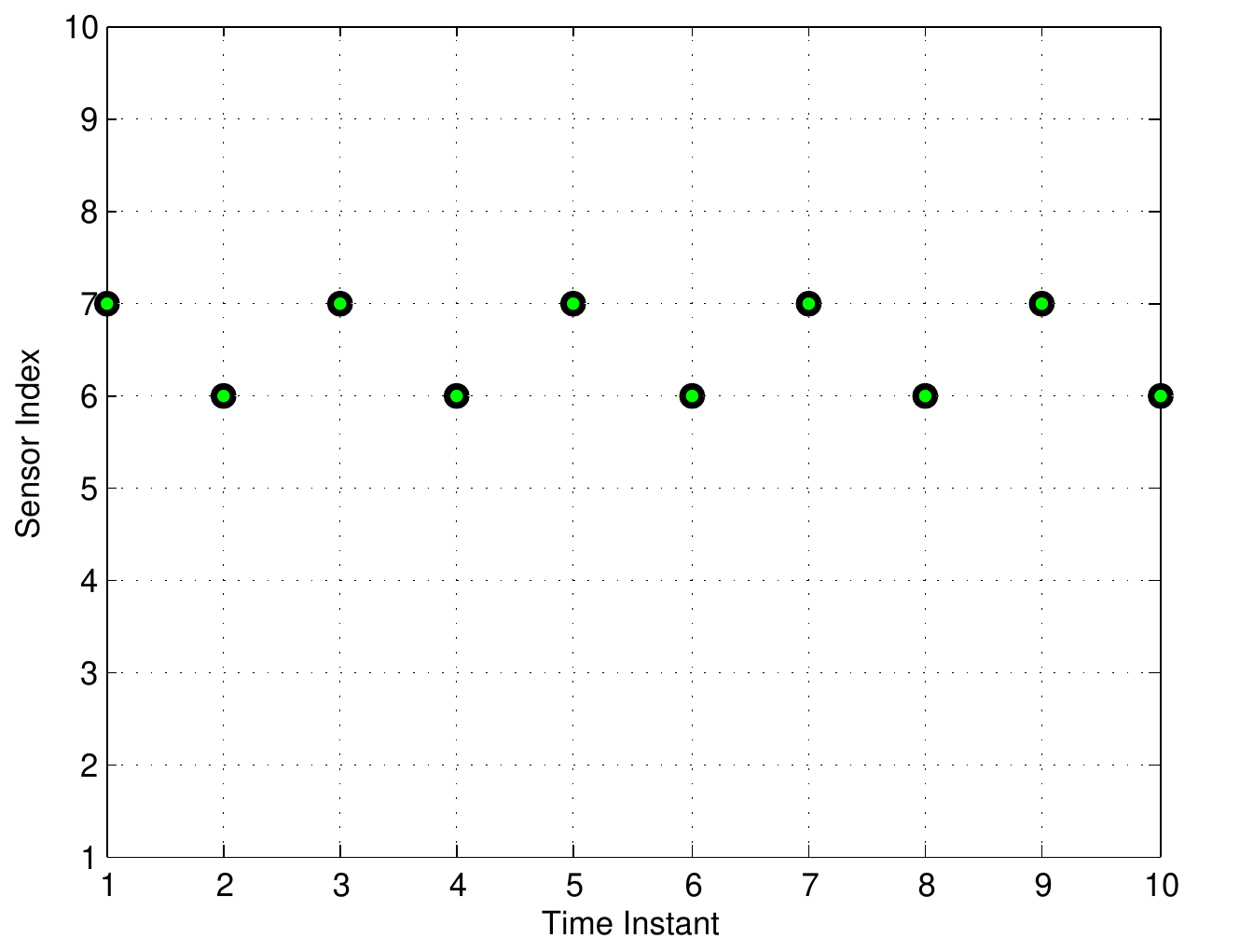} \\
(II-a) & (II-b) & (II-c) \\
\includegraphics[width=.3\textwidth,height=!]{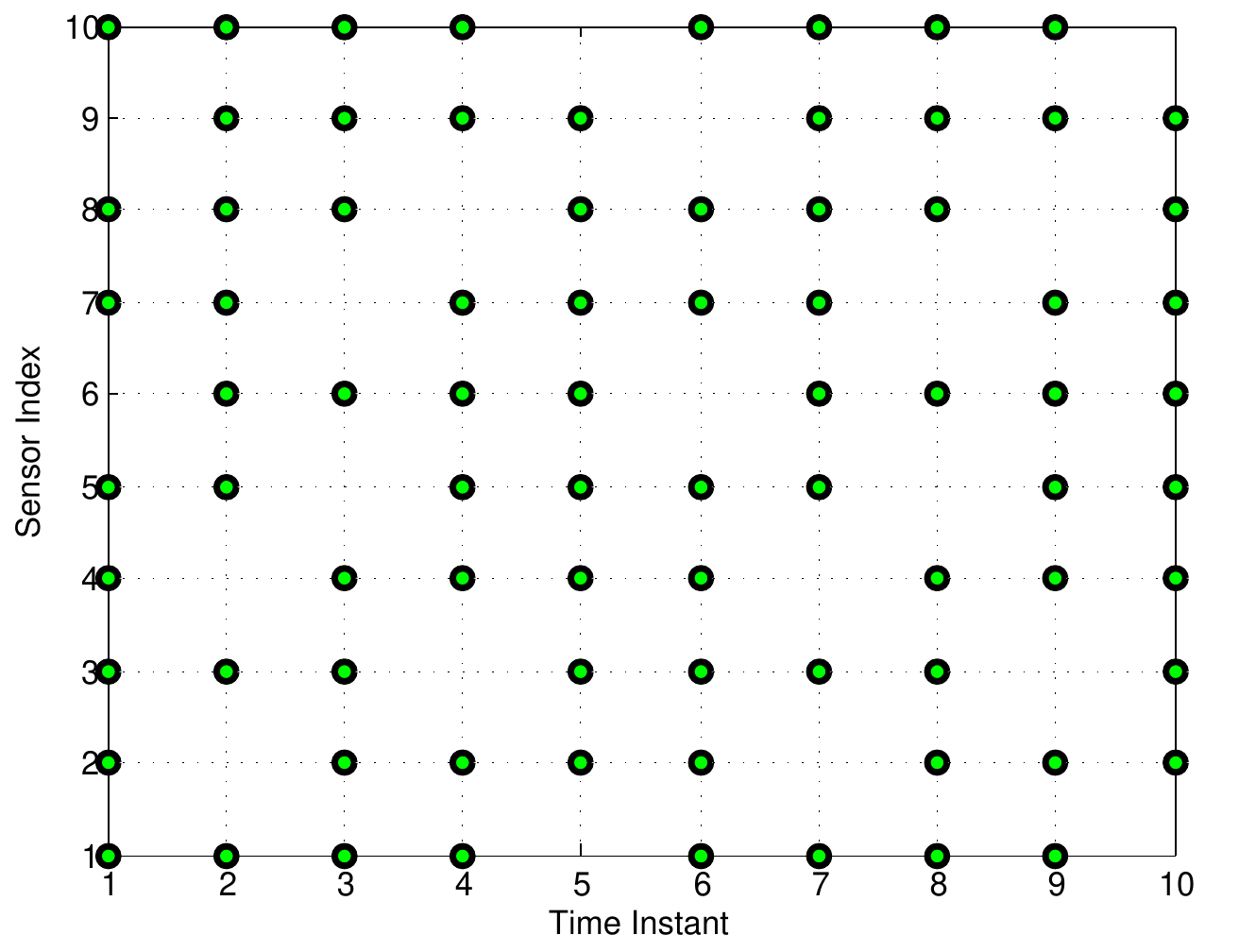} &
\includegraphics[width=.3\textwidth,height=!]{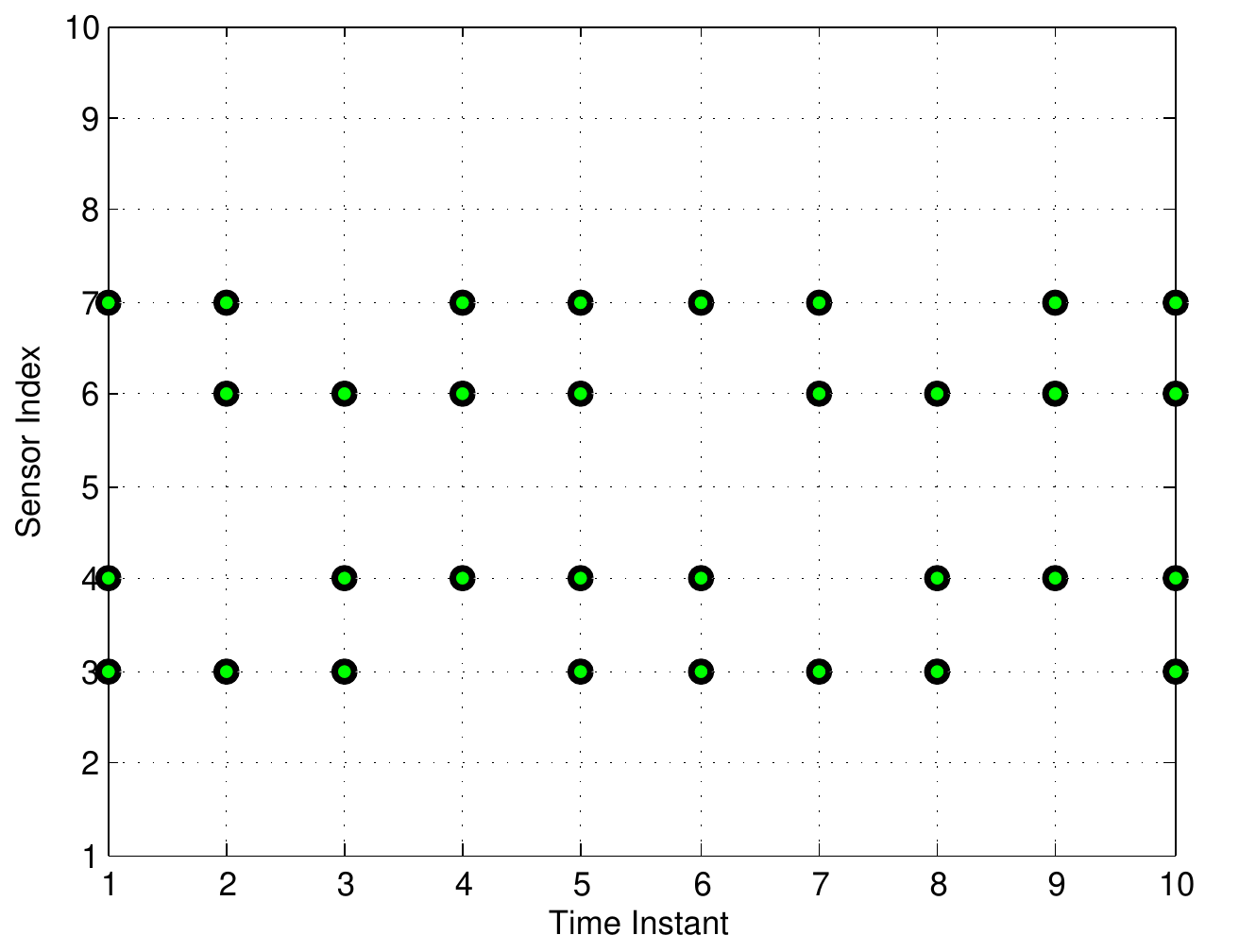} &
\includegraphics[width=.3\textwidth,height=!]{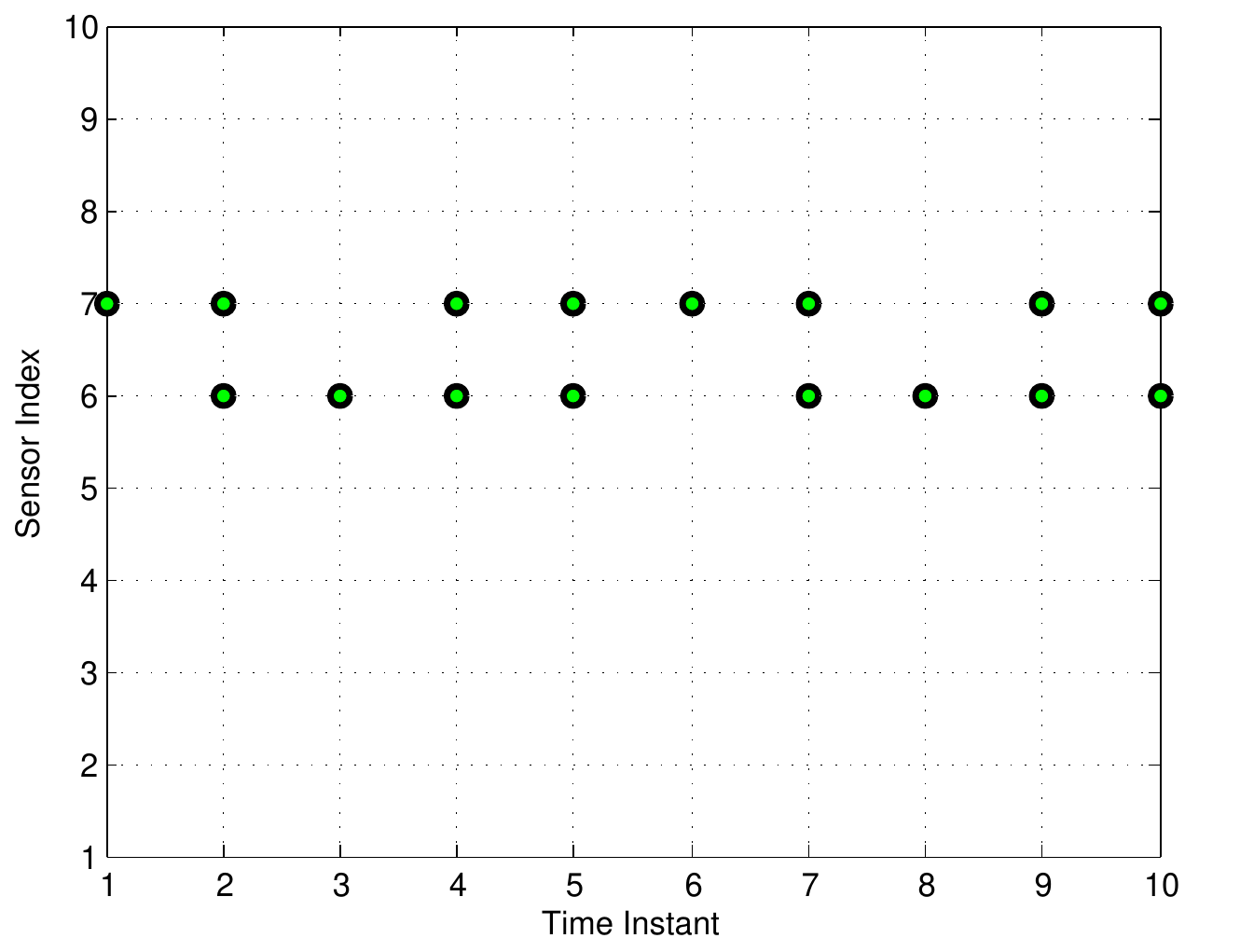} \\
(III-a) & (III-b) & (III-c) \\
\end{tabular}}
\caption{\footnotesize{
I- Sensor scheduling schemes with measurement frequency bound $\eta=1$:
	(I-a) $\gamma = 0$, 
	(I-b) $\gamma = 0.1$,
	(I-c) $\gamma = 0.15$.
II- Sensor scheduling schemes with measurement frequency bound $\eta=5$:
	(II-a) $\gamma = 0$, 
	(II-b) $\gamma = 0.1$,
	(II-c) $\gamma = 0.15$.
III- Sensor scheduling schemes with measurement frequency bound $\eta=8$:
	(III-a) $\gamma = 0$, 
	(III-b) $\gamma = 0.1$,
	(III-c) $\gamma = 0.15$.
}}
\label{fig: Sch}
\end{figure*}

In Fig.\,\ref{fig: Sch}, we use ADMM to obtain the sensor schedule over a time period of length $K=10$ for $\gamma \in \{0, 0.1, 0.15\}$ and $\eta \in \{ 1,5,8\}$; the subplots represent increasing values of $\gamma$ from left to right and increasing values of $\eta$ from top to bottom. In each subplot, the horizontal axis represents discrete time, the vertical axis represents sensor indices, and circles represent activated sensors. We also observe that sensors selected at two consecutive time instances tend to be spatially distant from each other. For example, at time instants $t = 1$, $2$, $3$, the active sensors are $6$, $9$, $4$, respectively.

In Figs.\,\ref{fig: Sch}-(I-a), (I-b), and (I-c), we assume $\eta = 1$ and vary the magnitude of the sparsity-promoting parameter $\gamma$. As seen in Fig.\,\ref{fig: Sch}-(I-a), for $\gamma = 0$ every sensor is selected exactly once over $K=10$ time steps. 
Figs.\,\ref{fig: Sch}-(I-b) and (I-c) demonstrate that fewer sensors are selected as $\gamma$ is further increased. This is to be expected, as the value of $\gamma$ in (\ref{eq: obj_1}) determines our emphasis on the column-cardinality of $\{L_k \}$. 

In Figs.\,\ref{fig: Sch}-(I-c), (II-c), and (III-c) for $\gamma = 0.15$ we compare the optimal time-periodic schedules for different values of the frequency bound $\eta=1,5,8$. Numerical results show that for $\gamma=0.15$ the $6$th and $7$th sensor are selected. 
To justify this selection, we note that these two sensors are located close to the center of the spatial region $\mathcal D$; see Fig.\,\ref{fig:sens_field}. Although we consider a random Gaussian field, the states at the boundary $\partial \mathcal D$ are forced to take the value zero and the states closest to the center of $\mathcal D$ are subject to the largest uncertainty. Therefore, from the perspective of entropy, the measurements taken from the sensors $6$ and $7$ are the most informative for the purpose of field estimation. As we increase $\eta$, we allow such informative sensors to be active more frequently.

Moreover, the sensor schedule in Fig.\,\ref{fig: Sch}-(II-c) verifies the optimality of the {\em uniform staggered sensing} schedule for two sensors, a sensing strategy whose optimality was proven in \cite{NVMn2005} and \cite[Proposition $5.2$]{Jiming2011}. In addition, although the periodicity of the sensor schedule was {\em a priori} fixed at the value $K=10$, as $\eta$ increases numerical results demonstrate repetitive patterns in the optimal sensor schedule. 
As seen in Figs.\,\ref{fig: Sch}-(II-c) and (III-c), for $\eta = 5$ and $\eta = 8$ the sensor schedule repeats itself five times over 10 time steps and two times over 10 time steps, respectively. This indicates that the value of the sensing period $K$ can be made smaller than $10$.

{
\subsection*{Comparison with existing methods}

In this subsection, 
we compare the performance of our approach with that of
methods proposed in \cite{Jiming2011,JEM11,TM2011} and an exhaustive search that enumerates all possible measurement sequences.
For tractability in the exhaustive search, we consider a small random field where $N=4$ and $\mathbf R = \mathbf I$. 
The values of other system parameters  $K$, $M$, $T$ and $\mathbf Q$ are specified in the following numerical examples.
Also we set $\gamma = 0$ to make our approach comparable to the existing methods in \cite{Jiming2011,JEM11,TM2011}. The ADMM stopping tolerance is chosen as $\epsilon = 10^{-3}$.

\begin{figure}[htb]
\centering
\includegraphics[width=.5\textwidth]{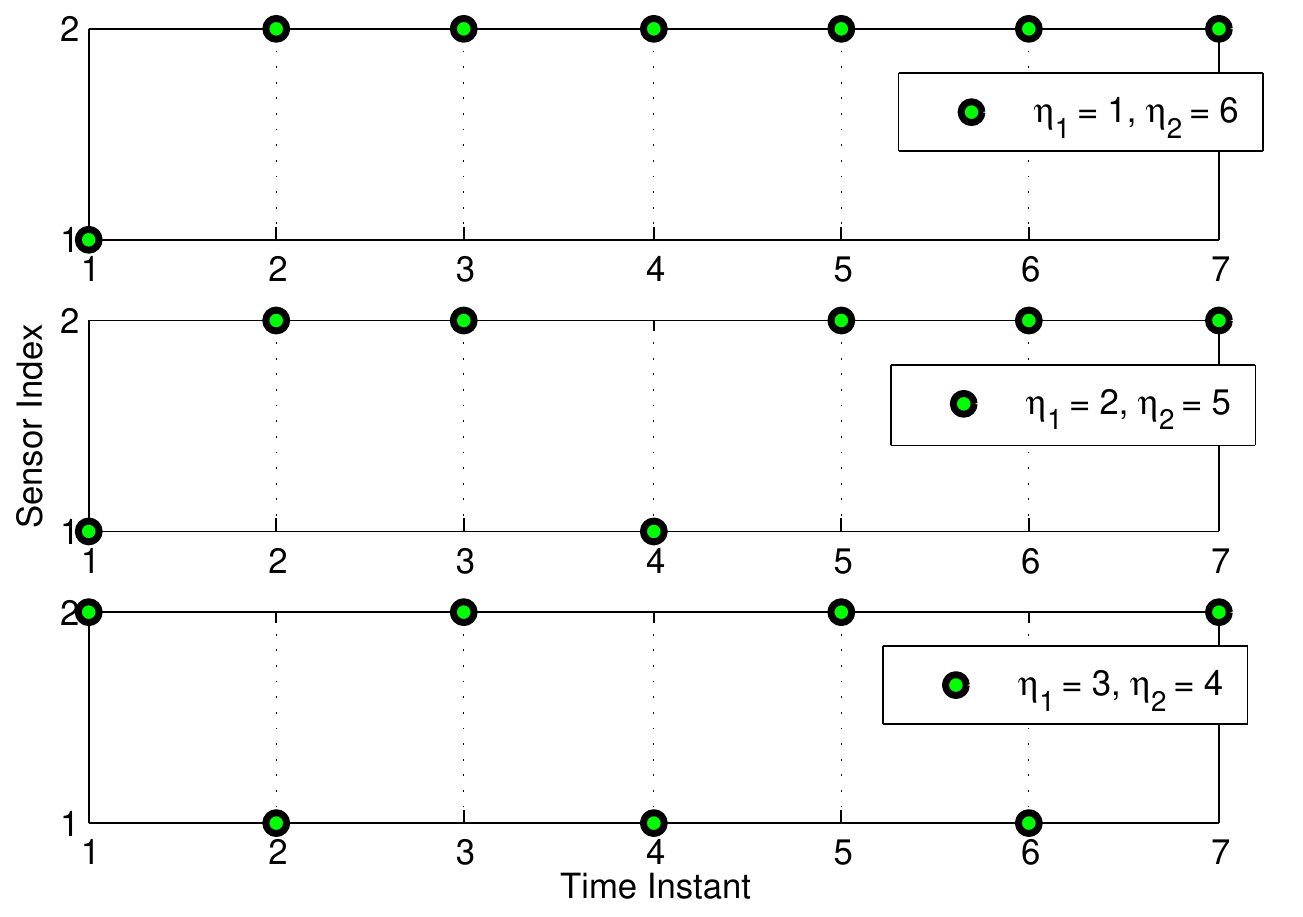}
\caption{\footnotesize{The periodic sensor schedule of two sensors with $K=7$.}}
\label{fig: sch_2s_SL}
\end{figure}

In Fig.\,\ref{fig: sch_2s_SL}, we study a numerical example stated in \cite{Jiming2011} and present the sensor schedules obtained from our approach with $K = 7$,  $M=2$, $T = 0.5$, $\mathbf Q = 0.25 \mathbf I$, $\eta_1 + \eta_2 = 7$ and $\eta_1 \in \{1 ,2 ,3\}$.
We observe that the sensor with smaller value of measurement frequency bound (in the present example, this corresponds to sensor one with $\eta_1 \in \{1 ,2 ,3\}$) is scheduled as temporally uniformly as possible; the resulting periodic sensor schedules are in agreement with those obtained in \cite[Prop. 5.2]{Jiming2011}. 

\begin{figure}[htb]
\centering
\includegraphics[width=.5\textwidth]{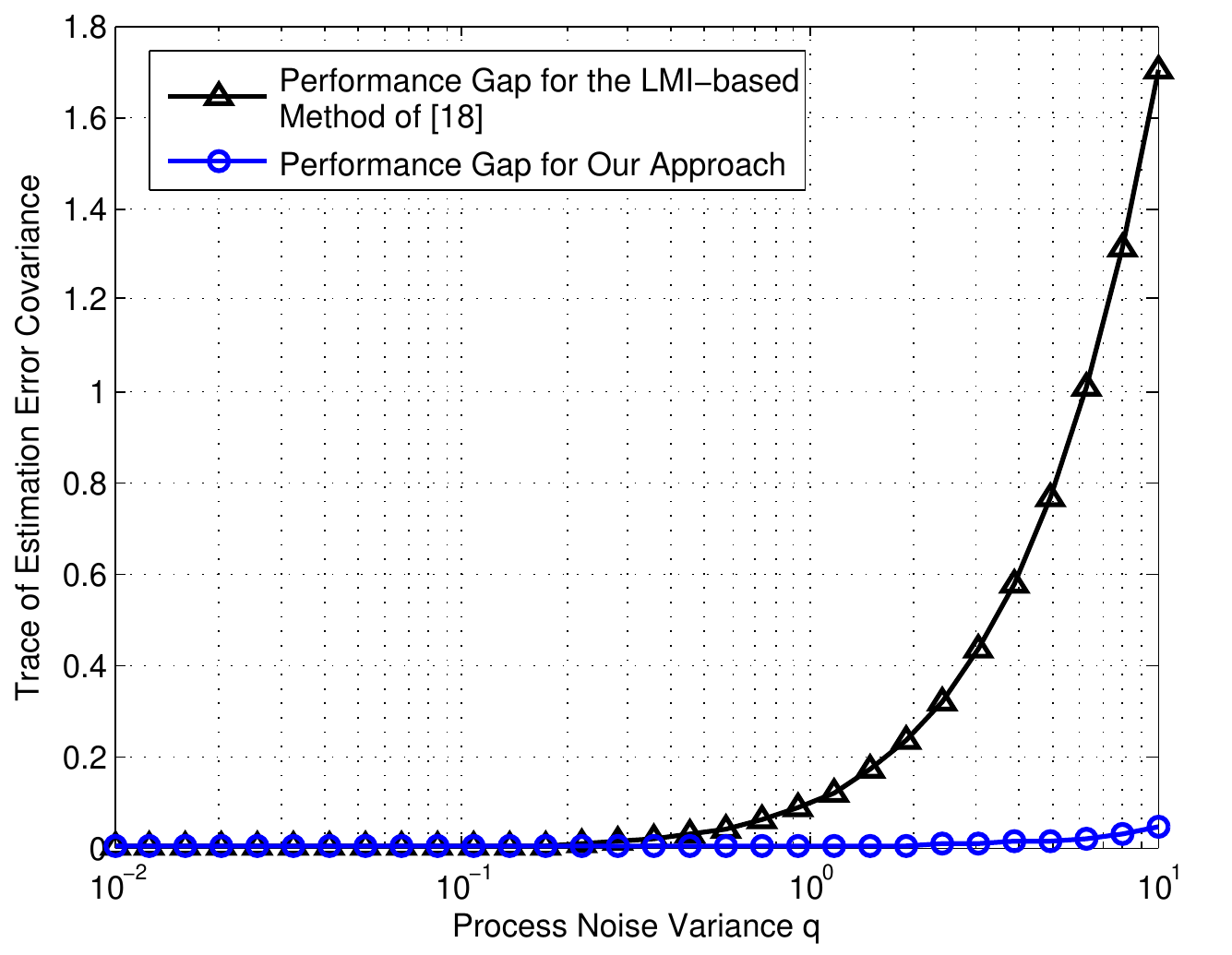}
\caption{\footnotesize{Performance gaps of the method in \cite{TM2011} and our
approach with respect to the optimal schedule.
}}
\label{fig: gap_Compell_AD_BF}
\end{figure}


In Fig.\,\ref{fig: gap_Compell_AD_BF}, we compare the performance of our approach with the periodic sensor scheduling method in \cite{TM2011}; each plot in Fig.\,\ref{fig: gap_Compell_AD_BF} represents the gap between the approach being considered and the globally optimal sensor schedule as a function of the process noise variance $q$, with $K = M = 4$,  $T = 0.5$, and $\mathbf Q = q \mathbf I$.
We recall that the periodic sensor scheduling problem in \cite{TM2011} is formulated under the assumption of negligible process noise, and is solved using linear matrix inequalities (LMIs). The assumption that process noise is negligible results in the insensitivity of the performance objective to the order in which sensors are activated. 
This assumption holds for deep space applications considered in [18], but is not a practical assumption in general.
Fig.\,\ref{fig: gap_Compell_AD_BF} demonstrates that for small values of $q$, our approach and that of \cite{TM2011} yield very similar performance.
However, for $q > 0.1$ our approach results in significant improvement in estimation performance. That is due to the fact that our optimization procedure takes into account the temporal ordering of sensor measurements.

In Fig.\,\ref{fig: Perf_comp_CKF}, we compare the performance of our proposed sensor scheduling approach and the periodic switching policy of \cite{JEM11}, where $M = 4$ and $Q = 0.25 \mathbf I$.
In Fig.\,\ref{fig: Perf_comp_CKF}-(a), each plot represents the gap between the approach being considered and the globally optimal sensor schedule for different values of the sampling interval $T$, where $K = 4$, and $T$ denotes the sampling time used to discretize the continuous-time system in (\ref{eq: heat_eq}). 
Since $K$ is the period in discrete time, the period length in continuous time is given by $\epsilon = KT$. 
Simulation results show that both of the sensor scheduling methods achieve the performance of the globally optimal sensor schedule as $T \to 0$. This is due to the fact that $\epsilon \to 0$ while $T \to 0$. And it has been shown in \cite{JEM11} that the best estimation performance is attained by using a periodic switching policy
as $\epsilon \to 0$ (and thus sensors are switched as fast as possible).
However, as $T$ increases, our approach outperforms the periodic switching policy significantly, which indicates that the sensor schedules obtained from the method of \cite{JEM11} are inappropriate for scheduling sensors for discrete-time systems with moderate sampling rates. 

\begin{figure}[htb]
\centerline{ 
\begin{tabular}{c}
\includegraphics[width=.5\textwidth,height=!]{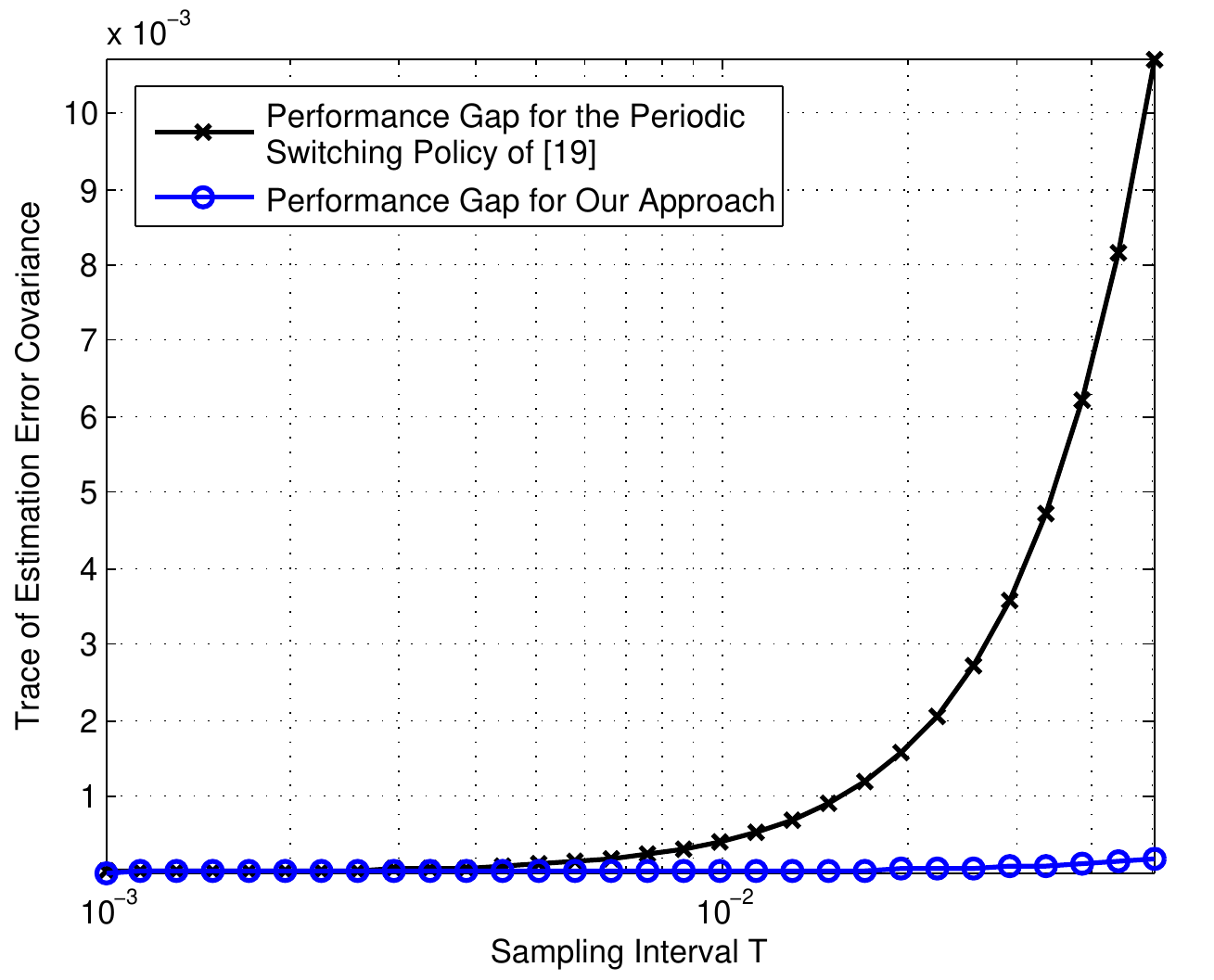}  \\
(a)\\
\includegraphics[width=.5\textwidth,height=!]{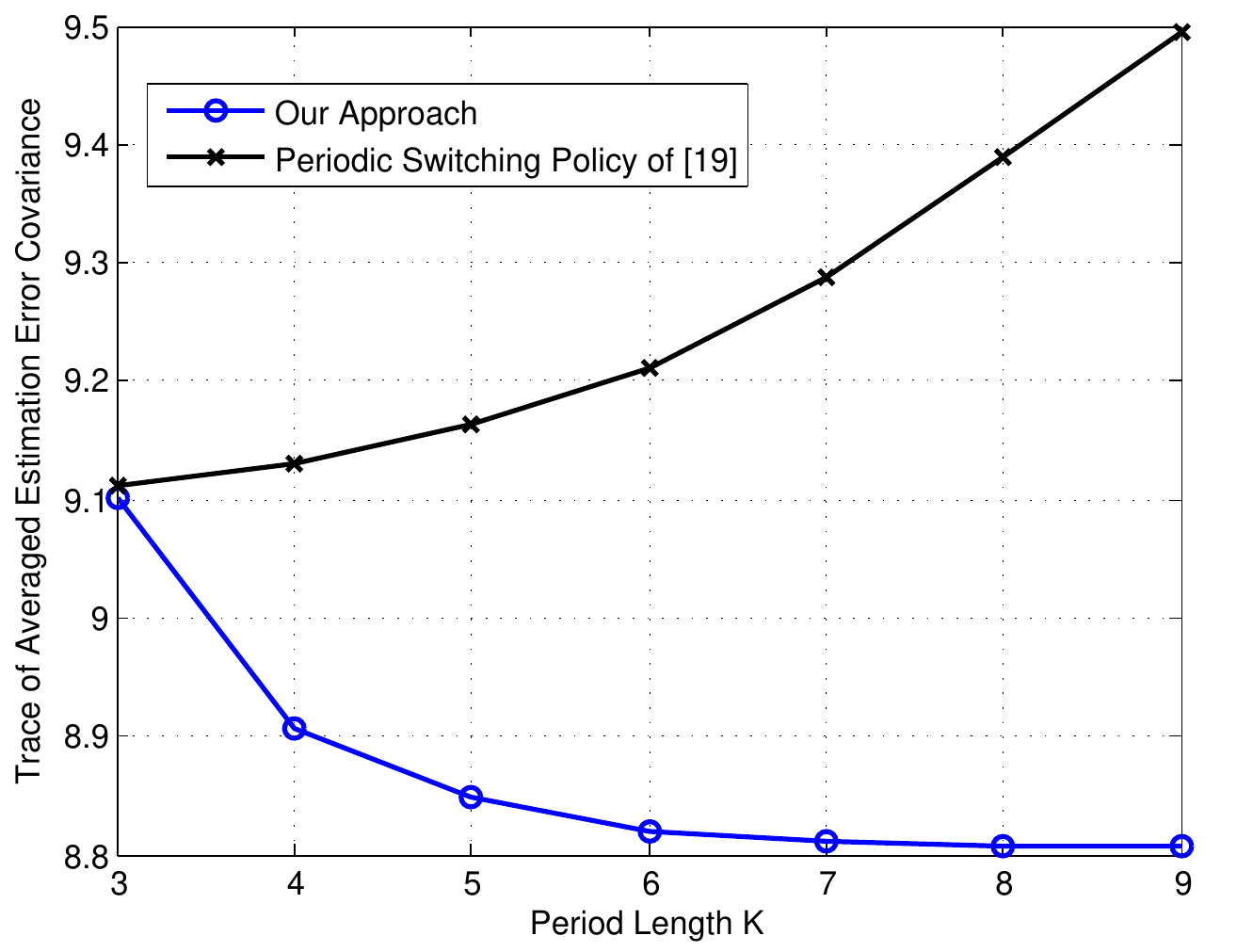} \\
(b) 
\end{tabular}}
\caption{\footnotesize{
Performance comparison with the periodic switching policy of \cite{JEM11}:
a) Performance gaps of the periodic switching policy of \cite{JEM11} and our
approach with respect to the optimal schedule;
b) Performance comparison between our approach and the periodic switching policy of \cite{JEM11} for different values of period lengths.
}}
\label{fig: Perf_comp_CKF}
\end{figure}

In Fig.\,\ref{fig: Perf_comp_CKF}-(b), we use the periodic switching policy of \cite{JEM11} to obtain the optimal sensor schedule and compare its performance with our approach for the 
fixed sampling interval $T = 0.5$ and different values of $K$.
Fig.\,\ref{fig: Perf_comp_CKF}-(b) demonstrates that the periodic switching policy of \cite{JEM11} {loses} optimality as $K$ increases.
This is not surprising, since the optimality of the periodic switching policy is only guaranteed as the period length goes to zero. Therefore, to schedule sensors on a discrete-time system with a moderate sampling interval, such as $T=0.5$ in the present example, our approach achieves better estimation performance than the periodic switching policy of \cite{JEM11}.

}


\section{Conclusion}
\label{sec:conclusion}

In this paper, we studied the problem of sensor scheduling for linear dynamical systems. We proposed an algorithm that determines optimal time-periodic sensor schedules. In order to strike a balance between estimation accuracy and the number of sensor activations, the optimization problem aims to minimize the trace of the estimation error covariance matrices while penalizing the number of nonzero columns of the estimator gains. We employed ADMM, which allows the  optimization problem to be decomposed into subproblems that can either be solved efficiently using iterative numerical methods or solved analytically to obtain exact solutions.
Our results showed that our approach outperforms previously available periodic sensor scheduling algorithms.

In this paper, we assumed that the period length of the periodic sensor schedule is fixed and is not an optimization variable. This leaves open the question of how to find the optimal period. Also, we characterized the sensor energy cost in terms of the number of times each sensor can be activated over a period. In future work, instead of a ``hard" constraint on the measurement frequency bound, we could consider more practical energy models to take into account the cost of repetitively selecting the informative sensors.
Furthermore, in order to reduce the computation burden of the fusion center, developing a decentralized architecture where ADMM can be carried out in a distributed way and by the sensors themselves is another future research direction.

\appendices

\section{Proof of Proposition 1}
\label{sec: append1}

The optimization problem (\ref{eq: L1_obj}) is equivalent to 
\begin{equation} \label{eq: L_obj_appendix}
\begin{array}{ll}
\hspace*{-0.3cm}\text{minimize}
&
\phi(\{\mathbf L_k\}) 
=  \displaystyle{\sum_{k=0}^{K-1}} \mathrm{tr}(\mathbf P_k) 
+ \displaystyle{\frac{\rho}{2}} \sum_{k=0}^{K-1} \mathrm{tr} [ (\mathbf L_k - \mathbf U_k)^T(\mathbf L_k - \mathbf U_k) ]
\\[0.15cm]
\hspace*{-0.3cm}\text{subject to}
&
\mathbf P_{k+1} =  (\mathbf A-\mathbf L_k{\mathbf C})\mathbf P_k(\mathbf A-\mathbf L_k{\mathbf C})^T
+\mathbf B \mathbf Q \mathbf B^T+ \mathbf L_k{ \mathbf R}\mathbf L_k^T.
\end{array}
\tag{$L$-$\Phi$}
\end{equation}
To find the necessary conditions for optimality, we 
find the gradient of $\phi$ and set $\nabla_{\mathbf L_k}\phi = 0$ for $k=0,1, \cdots, K-1$.

We begin by assuming an incremental change in the unknown variables $\{\mathbf L_k\}$ and finding the resulting incremental change to the value of the objective. Replacing $\mathbf L_k$ with $\mathbf L_k + \delta \mathbf L_k$ and $\phi$ with $\phi + \delta \phi$ in the objective function of (\ref{eq: L_obj_appendix}) and collecting first order variation terms on both sides, we obtain 
\begin{align*}
\delta \phi = & \sum_{k=0}^{K-1} \tr (\delta \mathbf P_k) 
+ \frac{\rho}{2} \sum_{k=0}^{K-1} \tr[ (\mathbf L_k - \mathbf U_k)^T \delta \mathbf L_k 
+ \delta \mathbf L_k^T ( \mathbf L_k - \mathbf U_k ) ].
\end{align*} 
We note that for $\delta \mathbf L_k$ to constitute a legitimate variation of $\mathbf L_k$, it has to satisfy the periodicity property $\delta \mathbf L_{k+K} = \delta \mathbf L_k$.
Similarly, replacing $\mathbf L_k$ with $\mathbf L_k + \delta \mathbf L_k$ and $\mathbf P_k$ with $\mathbf P_k + \delta \mathbf P_k$ in the constraint equation of (\ref{eq: L_obj_appendix}) and collecting first-order variation terms on both sides, we obtain
\begin{align*}
\delta \mathbf P_{k+1} = (\mathbf A- \mathbf L_k \mathbf C) \delta \mathbf P_k ( \mathbf A- \mathbf L_k \mathbf C)^T + \delta \mathbf M_k,
\end{align*} 
where
\begin{align*}
\delta \mathbf M_k =  - \delta \mathbf L_k \mathbf C \mathbf P_k (\mathbf A- \mathbf L_k \mathbf C)^T - (\mathbf A- \mathbf L_k \mathbf C) \mathbf P_k \mathbf C^T \delta \mathbf L_k^T  + \delta \mathbf L_k \mathbf R \mathbf L_k^T + \mathbf L_k \mathbf R \, \delta \mathbf L_k^T.
\end{align*}
The difficulty with finding the gradient of $\phi$ from the above equation is the dependence of $\delta \phi$ on $\delta \mathbf P_k$, with the dependence of $\delta \mathbf P_k$ on $\delta \mathbf L_k$ being through a Lyapunov recursion. In what follows, we aim to express $\sum_{k=0}^{K-1} \tr(\delta \mathbf P_k)$ in terms of $\{ \delta \mathbf L_k \}$.

It is easy to see that
\begin{equation*}
\begin{array}{lll}
\delta \mathbf P_k \hspace{-0.1in}  = \delta \mathbf M_{k-1} + \hspace{-0.05in} \displaystyle  \sum_{n=k-1}^{-\infty} 
(\mathbf A - \mathbf L_{k-1} \mathbf C) \cdots (\mathbf A-\mathbf L_n \mathbf C) 
\cdot \delta \mathbf M_{n-1} 
\cdot (\mathbf A - \mathbf L_n \mathbf C)^T \cdots (\mathbf A-\mathbf L_{k-1}\mathbf C)^T.
\end{array}
\end{equation*}
Taking the trace of both sides of the equation and summing over $k$, we have
\begin{align*}
\sum_{k=0}^{K-1} \tr (\delta \mathbf P_k) 
&= \sum_{k=0}^{K-1} \tr (\delta \mathbf M_{k-1}) 
 + \sum_{k=0}^{K-1} \sum_{n=k-1}^{-\infty} \tr[(\mathbf A - \mathbf L_n \mathbf C)^T \cdots (\mathbf A-\mathbf L_{k-1} \mathbf C)^T \\
& \hspace{1.8in}
(\mathbf A - \mathbf L_{k-1} \mathbf C) \cdots (\mathbf A - \mathbf L_n \mathbf C) \delta \mathbf M_{n-1}],
\end{align*} 
where we have used the property of the trace to change the order of the terms inside the square brackets. Now exploiting the periodicity properties $\mathbf L_{k+K}= \mathbf L_k$, $\delta \mathbf L_{k+K} = \delta \mathbf L_k$, $\mathbf P_{k+K} = \mathbf P_k$, which also imply the periodicity $\delta \mathbf M_{k+K} = \delta \mathbf M_k$ of $\{\delta  \mathbf M_k \}$, the double sum in the last equation above can be rewritten to give
\begin{align*}
\sum_{k=0}^{K-1} \hspace*{-0.03in}\tr (\delta \mathbf P_k)  
=\hspace*{-0.03in} \sum_{n=0}^{K-1} \hspace*{-0.03in} \tr \{ [ \mathbf I + \hspace*{-0.1in} \sum_{k=n+1}^{+\infty} \hspace*{-0.05in} (\mathbf A \hspace*{-0.03in} -\hspace*{-0.03in}  \mathbf L_n \mathbf C)^T \hspace*{-0.03in} \cdots \hspace*{-0.01in} (\mathbf A \hspace*{-0.03in} - \hspace*{-0.03in} \mathbf L_{k-1}\mathbf C)^T
\mathbf I (\mathbf A - \mathbf L_{k-1} \mathbf C) \cdots (\mathbf A - \mathbf L_n \mathbf C)  ] \delta \mathbf M_{n-1}\}.
\end{align*} 

To help with the simplification of the above sums, we define the new matrix variable $\mathbf V_n$ as 
\begin{align*}
\mathbf V_n &= \mathbf I + \sum_{k=n+1}^{+ \infty}(\mathbf A - \mathbf L_n \mathbf C)^T  \cdots (\mathbf A - \mathbf L_{k-1} \mathbf C)^T \mathbf I 
(\mathbf A - \mathbf L_{k-1} \mathbf C) \cdots (\mathbf A - \mathbf L_n \mathbf C).
\end{align*}
It can be seen that $\{ \mathbf V_n \}$ is periodic, $\mathbf V_{n+K} =\mathbf V_n$, and satisfies
\begin{align*}
\mathbf V_n = (\mathbf A - \mathbf L_n \mathbf C)^T \mathbf V_{n+1} (\mathbf A - \mathbf L_n \mathbf C) + \mathbf I.
\end{align*}
Returning to $\sum_{k=0}^{K-1} \tr (\delta \mathbf P_k)$ and using the definition of $\mathbf V_n$, we obtain
\begin{align*}
\sum_{k=0}^{K-1} \tr(\delta \mathbf P_k) = \sum_{n=0}^{K-1} \tr (\mathbf V_n \delta \mathbf M_{n-1}) = \sum_{n=0}^{K-1} \tr (\mathbf V_{n+1} \delta \mathbf M_n),
\end{align*}
where the last equality results from the periodicity of $\{ \mathbf V_n \}$ and $\{ \delta \mathbf M_n\}$. Recalling that $\delta \mathbf M_n$ can be written explicity in terms of $\{ \delta \mathbf L_k \}$, we have thus achieved our goal of expressing $\sum_{k=0}^{K-1} \tr (\delta \mathbf P_k)$ in terms of $\{ \delta \mathbf L_k \}$. We next carry out the last step required to find the gradient of $\phi$.

Replacing $\sum_{k=0}^{K-1} \tr (\delta \mathbf P_k)$ with $\sum_{k=0}^{K-1} \tr (\mathbf V_{k+1}\delta \mathbf M_k)$ in the expression for $\delta \phi$, and using the definition of $\delta \mathbf M_k$, we obtain
\begin{align*}
\delta \phi =  & \sum_{k=0}^{K-1} \tr (\mathbf V_{k+1}\delta \mathbf M_k) +  \frac{\rho}{2} \sum_{k=0}^{K-1} \tr[ (\mathbf L_k - \mathbf U_k)^T \delta \mathbf L_k + \delta \mathbf L_k^T ( \mathbf L_k - \mathbf U_k ) ] \nonumber \\
 = & 2 \hspace*{-0.03in}  \sum_{k=0}^{K-1} \hspace*{-0.04in}  \tr[-\mathbf C \mathbf P_k (\mathbf A - \mathbf L_k \mathbf C)^T \mathbf V_{k+1} \delta \mathbf L_k \hspace*{-0.02in} + \hspace*{-0.02in} \mathbf R \mathbf L_k^T \mathbf V_{k+1}\delta \mathbf L_k ]  + \rho \sum_{k=0}^{K-1} \tr[(\mathbf L_k - \mathbf U_k)^T \delta \mathbf L_k],
\end{align*}
where we have used the properties of the trace to arrive at the last equality. Thus
\begin{align*}
\nabla_{\mathbf L_k}\phi 
= [-2\mathbf C \mathbf P_k (\mathbf A - \mathbf L_k \mathbf C)^T \mathbf V_{k+1}\hspace*{-0.02in}  +  \hspace*{-0.02in} \mathbf R \mathbf L_k^T \mathbf V_{k+1} \hspace*{-0.02in} + \hspace*{-0.02in} \rho (\mathbf L_k - \mathbf U_k)^T ]^T.
\end{align*}
Setting $\nabla_{\mathbf L_k}\phi = \mathbf 0$ gives the necessary condition for optimality
\begin{align*}
\mathbf 0 = -2 \mathbf V_{k+1} (\mathbf A - \mathbf L_k \mathbf C)\mathbf P_k \mathbf C^T + 2 \mathbf V_{k+1} \mathbf L_k \mathbf R + \rho (\mathbf L_k - \mathbf U_k),
\end{align*}
where $\mathbf P_k$ and $\mathbf V_k$ satisfy the recursion euqations
\begin{align*}
\mathbf P_{k+1} & = (\mathbf A - \mathbf L_k \mathbf C) \mathbf P_k (\mathbf A - \mathbf L_k \mathbf C)^T + \mathbf B \mathbf Q \mathbf B^T + \mathbf L_k \mathbf R \mathbf L_k^T \\
\mathbf V_k & = (\mathbf A - \mathbf L_k\mathbf  C)^T \mathbf V_{k+1} (\mathbf A - \mathbf L_k \mathbf C) + \mathbf I.
\end{align*}
The proof is now complete. \hfill $\blacksquare$


\section{Proof of Proposition 3}
\label{sec: append3}


Problem (\ref{eq: subG2_ineq}) is equivalent to
\begin{equation*}
\begin{array}{cl}
\text{minimize}
&
\psi_m(\mathbfcal G_m)
:=
\displaystyle{\sum_{k=0}^{K-1}} \gamma \card\! \big(  \|\mathbf G_{k,m}\|_2 \big)
 + \displaystyle{\sum_{k=0}^{K-1}} \frac{\rho}{2} \| \mathbf G_{k,m}- \mathbf S_{k,m} \| _2^2
\\[0.15cm]
\text{subject to}
& \displaystyle{\sum_{k=0}^{K-1}} \card\! \big(  \| \mathbf G_{k,m} \|_2  \big) = q, ~~q \in \{0,\ldots,\eta \}
\end{array}
\end{equation*}
where $\mathbfcal G_m : = [\mathbf G_{0,m}, \cdots, \mathbf G_{K-1,m}] $. Similar to $\mathbfcal G_m$, we form the matrix ${\mathbfcal S}_m$ by picking out the $m$th column from each of the matrices in the set $\{\mathbf S_k \}$ and stacking them to obtain
$
\mathbfcal{S}_m
:=
[
     \mathbf S_{0,m}, \cdots, \mathbf S_{K-1,m}
].
$
We define $\kappa := \sum_{k=0}^{K-1} \card\! \big(  \| \mathbf S_{k,m}\|_2  \big)$, which
gives the column-cardinality of $\mathbfcal S_m$.

It can be shown that if $q = \kappa$ then the minimizer $\mathbfcal G_m^{\kappa}$ of problem (\ref{eq: subG2_ineq}) is $\mathbfcal S_m$, and $\psi_m(\mathbfcal G_m^{\kappa}) = \gamma \kappa$.
If $q > \kappa$, we have $\psi_m(\mathbfcal G_m) > \psi_m(\mathbfcal G_m^{\kappa})$ for arbitrary values of $\mathbfcal G_m \in F_q$ since $\psi_m(\mathbfcal G_m) = \gamma q + \frac{\rho}{2} \| \mathbfcal G_m - \mathbfcal S_m\|_F^2 $ which is greater than $\psi_m(\mathbfcal G_m^{\kappa}) = \gamma \kappa$. Therefore, the solution of (\ref{eq: subG1_ineq}) is only determined by solving the sequence of minimization problems (\ref{eq: subG2_ineq}) for $q = 0,1,\ldots, \mathrm{min}\{\eta,\kappa\}$
rather than ${q = 0,1, \cdots, \eta}$.


For a given $q \in \{0,1,  \mathrm{min}\{\eta,\kappa\} \}$, problem (\ref{eq: subG2_ineq}) can be written as
\begin{equation*}
\begin{array}{ll}
\displaystyle \minimize_{\mathcal{G}_m}
&
\displaystyle{\sum_{k=0}^{K-1}} \frac{\rho}{2}  ||\mathbf G_{k,m} - \mathbf S_{k,m} ||_2^2
\\[0.15cm]
\text{subject to}
& \displaystyle{\sum_{k=0}^{K-1}} \card\! \big(  \| \mathbf G_{k,m}\|_2 \big) = q.
\end{array}
\end{equation*}
For $q = 0$, the minimizer $\mathbfcal{G}_m^q$ of the optimization problem is $\mathbf 0$.
For $q \neq 0$, it was demonstrated in \cite[Appendix B]{FMM2011} that the solution is obtained by projecting the minimizer ($\mathbfcal G_m = \mathbfcal S_m$) of the objective function 
onto the constraint set $\sum_{k=0}^{K-1} \card\! \big( \|\mathbf G_{k,m}\|_2  \big) = q$. This gives
\begin{align}
\mathbf G_{k,m} = \left\{
  \begin{array}{l l}
    \mathbf S_{k,m} & \quad || \mathbf S_{k,m}||_2 \geq || [\mathbfcal S_m]_{q}||_2    \\
 \mathbf{0} & \quad  || \mathbf S_{k,m} ||_2 < || [\mathbfcal S_m]_{q} ||_2  \\
  \end{array}, \right.
\label{eq: sol_ini}
\end{align}
for $k = 0,1,\cdots, K-1$, where $[\mathbfcal S_m]_{q}$ is the $q$th largest column of $\mathbfcal S_m$ in the $2$-norm sense. The proof is now complete. \hfill $\blacksquare$


\section{Proof of Proposition 4}
\label{sec: append4}

According to Prop. \ref{prop_ineq}, substituting the minimizer $\mathbfcal G_m^q$ of problem (\ref{eq: subG2_ineq}) into its objective function yields
\begin{align*}
\psi_m(\mathbfcal G_m^q) &=\sum_{k = 0}^{K-1} \frac{\rho}{2} \|\mathbf G_{k,m} - \mathbf S_{k,m}  \|_2^2 + 
\sum_{k = 0}^{K-1} \gamma \mathrm{card}\left (\| \mathbf G_{k,m}\|_2 \right ) \nonumber \\
& = 
\sum_{\begin{subarray}{l} k = 0 \\
k \notin \chi_q \end{subarray}}^{K-1} \frac{\rho}{2} \| \mathbf S_{k,m}\|_2^2 + \gamma q, 
\end{align*}
where 
$\mathbf S_{k,m}$ denotes the $m$th column of $\mathbf S_k$, $\chi_q$ is a set that is composed by
indices of the first $q$ largest columns of $\mathbfcal{S}_m$ (refer to Appendix\,\ref{sec: append3}) in the $2$-norm sense, and $\chi_0 = \emptyset$.
We have
\begin{align}
\psi_m(\mathbfcal G_m^q)  - \psi_m(\mathbfcal G_m^{q+1})
= \frac{\rho}{2} \| [\mathbfcal S_m]_{q+1}\|_2^2 - \gamma
\label{eq: dif_Gq}
\end{align}
for $q = 0,1, \ldots, \mathrm{min}\{\eta,\kappa \}-1$, where $[\mathbfcal S_m]_{q+1}$ denotes the $(q+1)$th largest column of $\mathbfcal S_m$ in the $2$-norm sense.

Since $[\mathbfcal S_m]_{1} \geq [\mathbfcal S_m]_{2} \geq \cdots \geq [\mathbfcal S_m]_{\mathrm{min}\{\eta,\kappa \}}$, for $\gamma \in (\frac{\rho}{2} \| [\mathbfcal S_m]_{1}\|_2^2, \infty )$ equation (\ref{eq: dif_Gq}) yields $\psi_m(\mathbfcal G_m^0)  - \psi_m(\mathbfcal G_m^{1}) < 0$ and 
$\psi_m(\mathbfcal G_m^q)  - \psi_m(\mathbfcal G_m^{q+1}) < 0 $ for other $q \in \{1,\ldots, \mathrm{min}\{\eta,\kappa \}-1\}$. Therefore, the minimizer of (\ref{eq: subG1_ineq}) is given by
$\mathbfcal G_m^0$.
Similarly, for $\gamma \in (\frac{\rho}{2} \| [\mathbfcal S_m]_{q+1}\|_2^2, \frac{\rho}{2} \| [\mathbfcal S_m]_{q}\|_2^2]$ equation (\ref{eq: dif_Gq}) yields $\psi_m(\mathbfcal G_m^{l-1})  - \psi_m(\mathbfcal G_m^{l}) \geq 0$ for $l = 1,\ldots, q$, and $\psi_m(\mathbfcal G_m^{l})  - \psi_m(\mathbfcal G_m^{l+1}) < 0$ for $l = q,\ldots,\mathrm{min}\{\eta,\kappa \}-1$. Therefore, the minimizer of (\ref{eq: subG1_ineq}) is given by
$\mathbfcal G_m^q$.
Finally, we can write the solution of (\ref{eq: subG1_ineq}) in the form given in the statement of Prop. \ref{prop_G2}.
The proof is now complete. \hfill $\blacksquare$

\section*{Acknowledgment}

The authors acknowledge Dr. Fu Lin and Prof. Mihailo R. Jovanovi\'c for making available their Matlab codes for sparsity-promoting linear quadratic regulators (\verb"lqrsp.m", available at \texttt{http://www.ece.umn\\
.edu/users/mihailo/software/lqrsp/index.html}). These files facilitated the authors in implementing the algorithms presented in this work for optimal periodic sensor scheduling.

\ifCLASSOPTIONcaptionsoff
  \newpage
\fi



\bibliographystyle{IEEEbib}
\bibliography{Bibs/journal,Bibs/MakFarBib}  

\end{document}